\def\zem{$z_{\rm em}$}
\def\zabs{$z_{\rm abs}$}
\def\zphot{$z_{\rm phot}$}

\def\ha{H$\alpha$}

\def\oii{[O~{\sc ii}]}
\def\oiii{[O~{\sc iii}]}
\def\nii{[N~{\sc ii}]}

\def\hi{H~{\sc i}}
\def\hii{H~{\sc ii}}

\def\nhi{\mbox{$\sc N(\sc H~{\sc I})$}}
\def\lognhi{\mbox{$\log \sc N(\sc H~{\sc I})$}}

\def\civ{C~{\sc iv}}
\def\feii{Fe~{\sc ii}}

\def\mgi{Mg~{\sc i}}
\def\mgii{Mg~{\sc ii}}

\def\mnii{Mn~{\sc ii}}
\def\siii{Si~{\sc ii}}
\def\siiii{Si~{\sc iii}}

\def\tiii{Ti~{\sc ii}}

\def\alii{Al~{\sc ii}}
\def\aliii{Al~{\sc iii}}
\def\znii{Zn~{\sc ii}}
\def\crii{Cr~{\sc ii}}

\documentclass[useAMS,usenatbib,epsfig]{mn2e}
\usepackage{rotating}
\usepackage{float}

\title[Metallicity \& Geometry]{A SINFONI Integral Field Spectroscopy Survey for Galaxy Counterparts to Damped Lyman-$\alpha$ Systems - VI. Metallicity and Geometry as Gas Flow Probes\thanks{Based on observations collected during programmes ESO 078.A-0003, 092.A-0167 and 095.A-0338 at the European Southern Observatory with UVES, SINFONI and X-Shooter on the 8.2 m telescopes operated at the Paranal Observatory, Chile.} }
\author[C\'eline P\'eroux et al.] {C\'eline P\'eroux$^1$\thanks{e-mail: celine.peroux@gmail.com}, Samuel Quiret$^1$, Hadi Rahmani$^1$, Varsha P. Kulkarni$^2$, 
\newauthor
Benoit Epinat$^1$, Bruno Milliard$^1$, Lorrie A. Straka$^3$, Donald G. York$^4$,
\newauthor
Alireza Rahmati$^5$ \& Thierry Contini$^6$\\
$^1$ Aix Marseille Universit\'e, CNRS, LAM (Laboratoire d'Astrophysique de Marseille) UMR 7326, 13388, Marseille, France.  \\
$^2$ Dept. of Physics and Astronomy, Univ. of South Carolina, Columbia, SC 29208, USA.\\
$^3$ Sterrewacht Leiden, Leiden University, PO Box 9513, NL-2300 RA Leiden, the Netherlands.\\
$^4$ Dept. of Astronomy and Astrophysics and The Enrico Fermi Institute, University of Chicago, 5640 S. Ellis Ave, Chicago, IL 60637, USA.\\
$^5$ Institute for Computational Science, University of Zürich, Winterthurerstrasse 190, CH-8057 Zürich, Switzerland.\\
$^6$ IRAP, Institut de Recherche en Astrophysique et Plan\'etologie, CNRS, 14, avenue Edouard Belin, F-31400 Toulouse, France \\
   and Universit\'e de Toulouse, UPS-OMP, Toulouse, France.\\
}

\begin{document}

\date{Accepted 2016 January 01. Received 2015 December 21; in original form 2015 August 04}

\pagerange{\pageref{firstpage}--\pageref{lastpage}} \pubyear{2002}

\maketitle

\label{firstpage}

\begin{abstract}
The use of background quasars provides a powerful tool to probe the cool gas in the circum-galactic medium of foreground galaxies. Here, we present new observations with SINFONI and X-Shooter of absorbing-galaxy candidates at z=0.7--1. 
We report the detection with both instruments of the \ha\ emission line of one sub-DLA at \zabs=0.94187 with \lognhi=19.38$^{+0.10}_{-0.15}$ towards SDSS J002133.27+004300.9. We estimate the star formation rate: SFR=3.6$\pm$2.2 M$_{\odot}$/yr in that system. A detailed kinematic study indicates 
a dynamical mass M$_{\rm dyn}$=10$^{9.9\pm0.4}$ M$_{\odot}$ and a halo mass M$_{\rm halo}$=10$^{11.9\pm0.5}$ M$_{\odot}$. In addition, we report the \oii\ detection with X-Shooter of  another DLA at \zabs=0.7402 with \lognhi=20.4$\pm$0.1 toward Q0052$+$0041 and an estimated SFR of 5.3$\pm$0.7 M$_{\odot}$/yr. Three other objects are detected in the continuum with X-Shooter but the nature and redshift of two of these objects are unconstrained due to the absence of emission lines, while the third object might be at the redshift of the quasar. We use the objects detected in our whole \nhi-selected SINFONI survey to compute the metallicity difference between the galaxy and the absorbing gas, $\delta_{HI}(X)$, where a positive (negative) value indicates infall (outflow). We compare this quantity with the quasar line of sight alignment with the galaxy's major (minor) axis, another tracer of infall (outflow). We find that these quantities do not correlate as expected from simple assumptions. Additional observations are necessary to relate these two independent probes of gas flows around galaxies.
\end{abstract}
\begin{keywords}
Galaxies: formation -- galaxies: evolution -- galaxies: abundances -- galaxies: ISM -- quasars: absorption lines -- intergalactic medium
\end{keywords}

\section{Introduction}

\begin{table*}
\begin{center}
\caption{{\bf Targets List.} Summary of some of the properties of the quasar field, absorber and absorbing-galaxy. The magnitudes are AB magnitudes and the angular separations are refered to as "b".}
\label{t:targets}
\begin{tabular}{ccccccccc}
\hline\hline
Quasar{\bf $^a$}  		  &Coordinates$^a$ &V Mag  &z$_{\rm quasar}$&\zabs  &\nhi &b &b &Inst.\\
&&&&&&["] &[kpc] &\\
\hline
SDSS J002133.27+004300.9         &00 21 33.28+00 43 00.99   &18.2 	&1.244 &0.9420	&19.38$^{+0.10}_{-0.15}$ &10.8	&86	&SINFO/XSH/UVES\\
Q0052+0041                       &00 51 30.49+00 41 50.06   &18.5	&1.19  &0.7402	&20.4$\pm$0.1  &3.3	 &24	&SINFO/XSH\\
SDSS J011615.51-004334.7         &01 16 15.52-00 43 34.747  &18.7  	&1.275 &0.9127	&19.95$^{+0.05}_{-0.11}$ &8.1	 &64	&SINFO/XSH\\
J0138-0005                        &01 38 25.54-00 05 34.52   &18.8	&1.340 &0.7820	&19.81$^{+0.06}_{-0.11}$ &6.5	 &49	&SINFO\\
J045647.17+040052.9		 &04 56 47.18+04 00 52.94   &16.5	&1.345 &0.8596	&20.75$\pm$0.03 &0.8	 &14	&SINFO/XSH\\
Q0826-2230                       &08 26 01.58-22 30 27.20   &16.2	&0.910 &0.9110	&19.04$\pm$0.04 &5.0	 &40	&SINFO/XSH\\
SDSS J122836.8+101841.7          &12 28 36.88+10 18 41.92   &18.5	&2.306 &0.9376	&19.41$^{+0.12}_{-0.18}$ &4.6	 &37	&XSH\\
SDSS J152102.00-000903.1         &15 21 02.01-00 09 03.21   &18.5	&2.032 &0.9510	&19.40$^{+0.08}_{-0.14}$ &8.5	 &68	&SINFO/XSH\\     
\hline\hline 				       			 	 
\end{tabular}			       			 	 
\end{center}			       			 	 
\begin{minipage}{180mm}
{\bf Note:} \\
{\bf $^a$} SIMBAD coordinates unless the quasars is part of SDSS, in which case SDSS names are provided.\\
\end{minipage}
\end{table*}

While the large-scale intergalactic medium (IGM) and small-scale processes within the galaxies are now better understood, the next challenge to probe the history of baryons is to focus on the circum-galactic medium (CGM), a gas-rich region which lies around galaxies but inside their dark matter halos, at scales between 10-20kpc (galaxy's radius) and 100-200kpc (dark matter halo virial radius). In particular, the interactions between gas inflows (imposed by large-scale dark matter structures from IGM reservoirs) and outflows (launched by star-forming regions in galaxies and active supermassive black holes) are of paramount importance.

Outflows are commonly probed by the presence of interstellar absorption lines from cool gas superimposed on the stellar continuum which are blue-shifted by hundreds of km/s relative to the background galaxies' systemic velocities \citep{shapley03, steidel10}. Strong \mgii\ absorbers in particular \citep{tremonti07,martin09,martin12,schroetter15} have been observed to extend out to 100 kpc along the galaxies' minor axes \citep{bordoloi11}. Outflows are ubiquitous in galaxies at various redshifts \citep[e.g.][]{martin05,rupke05,tremonti07,weiner09,pettini01,steidel01,veilleux05}. Interestingly, the circum-galactic gas has also been probed in emission by \citet{steidel11}  who stacked narrow-band images of z$\sim$2-3 galaxies and reveal diffuse Ly$\alpha$ haloes extending to $\sim$80 kpc. However, given the unknown ionisation state and number of phases in the gas, it is at present very difficult to measure the mass in these outflows \citep[e.g.][]{genzel10, martin13}. Moreover, galaxies are believed to interact with the IGM by filling it with ionising photons and by injecting heavy elements formed in stars and supernovae through these supersonic galactic winds. Indeed, observations of the IGM indicate significant quantities of metals at all redshifts \citep{pettini03, ryanweber09, tumlinson11, cooksey13, werk13, shull14a}. The presence of these metals is interpreted as a signature of strong galactic outflows in various models \citep{aguirre01, oppenheimer06, pieri07}.

While observational evidence for outflows is growing, direct probes of infall are notoriously more difficult to gather. Nevertheless, cool gas inflows have recently been detected in a few objects \citep{sato09, martin12, bouche13,diamond15}. Moreover, accretion is required to explain some of the basic observed properties of galaxies including the gas-phase metallicity \citep[e.g.][]{erb06a} and the cosmological evolution of neutral gas mass \citep{zafar13b}. From a theoretical viewpoint, baryonic infall along the cosmic web onto dark halos is relatively well understood in $\Lambda$-CDM models. 

The spectra of background quasars provide a powerful tool to probe the cool gas in the CGM. The quasar absorbers, which trace the diffuse gas environment surrounding galaxies, are used to study the CGM of galaxies \citep{stewart11, stinson12, rudie12}. However, studying the stellar content of these systems has proved to be very challenging. It is critical to select the rare quasar-galaxy pairs. In recent studies, we have used sight-lines selected by their large neutral gas content, Damped Lyman--$\alpha$: DLA ($\lognhi>$20.3) and sub-DLA (19.0$<\lognhi<$20.3) systems at z$\sim$1 and z$\sim$2 \citep{peroux11a, peroux11b, peroux12}. Similarly, a sample of \mgii\ absorbers was used to study winds and infall near galaxies at z$\sim$1 and z$\sim$2 \citep{bouche07, bouche13, schroetter15}, up to a redshift close to the peak of cosmic star formation activity. In these studies, we made use of the advantages afforded by the 3D spectroscopy at near-infrared wavelengths made possible by the SINFONI instrument on the European Very Large Telescopes (VLT) aided with the adaptive optics (AO) system to search for and study absorbing-galaxies.

The present paper is structured as follows. A summary of observational details and data reduction steps are provided in Section 2. 
In section 3, we report the detections of the absorbing-galaxies with these new data. Finally, the results for the full \nhi-selected sample are presented in Section 4. Throughout this paper, we assume a cosmology with H$_0$=71 km/s/Mpc, $\Omega_M$=0.27 and $\Omega_{\rm \Lambda}$=0.73.

\section{The Data}

\subsection{Target Selection}

The target selection in this study differs from previous work by our group \citep{bouche07, peroux11a, peroux13, schroetter15}. Previously, we performed blind search for emission at the redshift of the quasar absorber. In this case, we select absorbing-galaxy candidates in the quasar field with a known sky position. The candidates are identified based on either a proximity argument (i.e. small impact parameter to the quasar line-of-sight), a photometric redshift, or a spectroscopic redshift. We emphasize that this is {\it not} a blind search, but a set of eight z$\sim$1 quasar/absorbing-galaxies matched pairs identified by other studies. 
We collect from the available literature absorbers with known column density in neutral hydrogen, \nhi, 
with galaxy counterparts identified from either broad-band imaging \citep{lebrun97, meiring11},
with photometric redshifts \citep{ rao11} or with redshifts confirmed by low-resolution spectroscopy \citep{lacey03}. We carefully select objects with redshift so that the $\lambda_{\rm obs}$ of \ha\ emission falls away from OH sky lines. 
The neutral gas column density for these systems has been measured from HST/FOS or HST/STIS spectra. We originally requested nine targets to be observed with SINFONI and eight with X-Shooter. A ninth object (J0138$-$005) has a UVES quasar spectrum available in the archives, so no further observations are needed. In the end, the SINFONI and X-Shooter targets do not exactly overlap due to the fact that our programme is not complete: one field was observed with X-Shooter but not with SINFONI (SDSS J122836.8+101841.7). Finally, One field was not observed with either SINFONI or X-Shooter. Table~\ref{t:targets} provides a summary of some of the properties of these targets, including the coordinates of the quasar in the field, quasar magnitude and redshift, absorber redshift, \nhi\ column density, angular separation between the quasar and the absorbing-galaxy for each of the objects and which instrument the field is observed with.

\subsection{Observations and Data Reduction}

\subsubsection{SINFONI}

\begin{table*}
\begin{center}
\caption{{\bf Journal of high-resolution SINFONI observations.} The data are J-band and 8 arcsec $\times$ 8 arcsec field of view, corresponding to a 0.25-arcsec pixel scale.}
\label{t:JoO_SINFO}
\begin{tabular}{ccccccccc}
\hline\hline
Quasar 		  &Observing Date &T$_{\rm exp}$[sec]$\times $N$_{\rm exp}$$^a$	&AO$^b$ &PSF$^c$ \\
&&&&["]\\
\hline
SDSS J002133.27+004300.9         &2013 Oct 7		  &(600$\times$4)$\times$(0+1)	&no AO	   &0.75\\
Q0052+0041                       &2013 Nov 7		  &(600$\times$4)$\times$(2+0)	&no AO	   &0.95\\
SDSS J011615.51-004334.7         &2013 Oct 5/6	  	  &(600$\times$4)$\times$(1+1)	&no AO	   &-\\
J0138-0005                        &2013 Oct 30/Nov 9    &(600$\times$4)$\times$(2+0) 	&no AO	   &0.75\\
J045647.17+040052.9		 &2013 Dec 27     	  &(600$\times$4)$\times$(0+2) 	&NGS       &0.90\\
Q0826-2230                       &2013 Dec 14/Dec 30   &(600$\times$4)$\times$(2+0)	&NGS   	   &0.70\\
SDSS J152102.00-000903.1         &2014 Mar 22	  &(600$\times$4)$\times$(1+0)		&no AO	   &-\\         
\hline\hline 				       			 	 
\end{tabular}			       			 	 
\end{center}			       			 	 
\begin{minipage}{180mm}
{\bf Note:} \\
{\bf $^a$} The two numbers for N$_{\rm exp}$ in brackets refer to exposures classified as "completed" and "executed" (i.e. not within the user specifications in ESO terminology), respectively.\\  
{\bf $^b$} No AO: no Adaptive Optics, natural seeing. NGS: the quasar is used as a Natural Guide Star for Adaptive Optics.\\
{\bf $^c$} The seeing is measured from the quasar in the data cube. For two fields, the quasar is not covered. We note that during the exposure of SDSS J002133.27+004300.9, the seeing degraded to 0.95" so that the OB is not considered complete by ESO.\\ 
\end{minipage}
\end{table*}			       			 	 

Unlike in our previous studies with  SINFONI \citep{peroux11a, peroux12}, these new observations are centered on the absorbing-galaxy with known sky position. The observations were carried out in service mode (under programme ESO 92.A-0167 A) at the European Southern Observatory with SINFONI on the 8.2 m YEPUN telescope. The redshifted \ha\ line lies in the $J$-band for these targets at z$\sim$1. The data are 8 arcsec $\times$ 8 arcsec field of view, corresponding to a 0.25-arcsec pixel scale. In order to avoid sky exposures, we nod around the target. The resulting cubes (a mosaic of the 8"$\times$8" SINFONI field-of-view resulting in a $\sim$11"$\times$11" effective field-of-view) are centered on the absorbing-galaxy with known sky positions. For two fields where the quasar itself is bright enough, we use it as a natural guide star (NGS) for adaptive optics (AO) in order to improve the spatial resolution. For the other fields the data are taken with natural seeing (no AO). The seeing is measured from the quasar in the data cube if present. For two fields, the quasar is not covered. The resulting PSF is fairly poor with FWHM ranging from 0.70 to 0.95 arcsec. In fact, a number of {\rm Observing Blocks (OBs)} are classified as "executed", i.e. the observations are taken but not validated as up to ESO standards. The $J$-band grism provides a spectral resolution of around R$\sim$2000.  A journal of observations summarising the target properties and experimental set-up is presented in Table~\ref{t:JoO_SINFO}. The table provides the observing date, exposure time, adaptive optics system used and the resulting PSF of the combined data.

The data are reduced with the latest version of the ESO SINFONI pipeline (version 2.5.2) and custom routines. The latter are used to correct the raw cubes for bad detector columns and to remove cosmic rays by applying the Laplacian edge technique of \citet{vandokkum01}. Master bias and flat images based on calibration cubes taken closest in time to the science frames are used to correct each data cube. Bias and flat-field corrections are done within the ESO pipeline. The OH line suppression and sky subtraction are accomplished with additional purpose-developed codes. Within one OB, the science frames are pair-subtracted with an ON-OFF pattern to eliminate variation in the near-infrared sky background. The wavelength calibration is based on the Ar lamp and is accurate to about $\sim$ 30 km/s in the J-band, i.e. comparable with the calculated heliocentric correction (of the order of 10-30 km/s). For each set of observations, a flux standard star is observed at approximately the same time and at similar airmass as used for the target fields. The flux standard star data are reduced in the same way as the science data. These standard stars are then used for flux calibration by fitting a black body spectrum to the O/B stars or a power law to the cool stars (T$<$10,000K) and normalising them to the 2MASS magnitudes. These spectra are also used to remove atmospheric absorption features from the science cubes. When a quasar is included in the cube, the resulting flux calibration is compared with the quasar 2MASS magnitudes in order to estimate the flux uncertainties. The different observations from the independent Observing Blocks are then combined spatially using the position of the quasar in each frame or the PSF calibrator (a bright PSF star), resulting in an average co-added cube for each target.

\subsubsection{X-Shooter}

\begin{table*}
\begin{center}
\caption{{\bf Journal of X-Shooter Observations.} The medium-resolution X-Shooter spectrograph covers the full wavelength range from 300 nm to 2.5 $\mu$m. Our observing strategy consists in aligning the slit on both the bright background quasar and the faint absorbing-galaxy for which the exact sky positions are known from previous observations, except in one case where the angular separation is larger than the slit length (SDSS J002133.27+004300.9). 
}
\label{t:JoO_XSH}
\begin{tabular}{ccccccrlccc}
\hline\hline
Quasar		   &Observing Date &T$_{\rm exp}\times$ N$_{\rm exp}$ &Quasar? &Nodding Mode &Nodding Length \\
&&[sec]&&&["]\\
\hline
SDSS J002133.27+004300.9         &2013 Nov 17/Dec 16+2015 Jul 2    &(1200$\times$2+1200$\times$2)  &n &AutoNod	  &5\\	
Q0052+0041                       &2013 Oct 13-21-23/Nov 21  	     &(1440$\times$2+1200$\times$1)  &y  &GenericOffset &6\\	
SDSS J011615.51-004334.7         &2013 Oct 13/Nov 21	    	     &(1200$\times$2)	       &y  &GenericOffset &2\\	
J045647.17+040052.9	         &2013 Dec 5-11-21-22-24	     &(1440$\times$6+1200$\times$1)  &y &AutoNod	  &5\\	
Q0826-2230              	 &2013 Dec 12-15-24/Jan 8-19 	     &(1200$\times$3)	       &y &GenericOffset &3\\	
SDSS J122836.8+101841.7          &2014 Feb 13-22/Mar 7-12   	     &(1200$\times$3)	       &y &GenericOffset &3\\	
SDSS J152102.00-000903.1         &2014 Mar 16			     &(1200$\times$1)	       &y  &GenericOffset &2\\	
\hline\hline 				       			 	 
\end{tabular}			       			 	 
\end{center}			       			 	 
\begin{minipage}{180mm}
\end{minipage}
\end{table*}

Seven of the nine targets initially planned are observed with X-Shooter. The observations are carried out in service mode at the European Southern Observatory with X-Shooter \citep{vernet11} on the 8.2 m KUEYEN telescope. Most of the observations are taken under programme ESO 92.A-0167 B, while the object we detect, SDSS J002133.27+004300.9, is re-observed in ESO 095.A-0338 B. The medium-resolution X-Shooter spectrograph covers the full wavelength range from 300 nm to 2.5 $\mu$m thanks to the simultaneous use of three spectroscopic arms (UVB, VIS and NIR). Our observing strategy consists of aligning the slit on both the bright background quasar and the faint absorbing-galaxy for which the exact sky positions are known from previous publications. In one case (SDSS J002133.27+004300.9), the angular separation is larger than the slit length (11") plus nodding length, so that only the absorbing-galaxy is observed. We used the long-slit mode with slit width of 1.0" for UVB and 0.9" for the VIS and NIR arms. With these settings, the expected spectral resolution is 59 km/s (UVB), 34 km/s (VIS) and 53 km/s (NIR) respectively. Our actual resolution is only slightly higher than these estimates because the typical seeing (0.8") is only slightly smaller that the slit widths. To optimise the sky subtraction in the NIR, the nodding mode is used following an ABBA scheme. The nodding length is chosen to avoid both the quasar trace and quasar counter images falling at the absorbing-galaxy position. As a result, the "AutoNod" mode is used when a nodding length of 5" is suitable and fine-tuned otherwise using the "GenericOffset" mode. The total exposure time is divided in Observing Blocks of 1200 or 1440 seconds each. A journal of observations summarising the target properties and experimental set-up is presented in Table~\ref{t:JoO_XSH}. The table provides the observing date, exposure times, whether the quasar is covered by the slit or not and the nodding mode for each of the fields. 

The data are reduced with the latest version (2.6.0) of the ESO X-Shooter pipeline first described in \citet{goldoni06} and additional external routines for the extraction of the 1D spectra and their combination. Master bias and flat images based on data taken closest in time to the science frames are used to correct each raw spectrum. Bias and flat-field correction are part of the ESO pipeline. Cosmic rays are removed by applying the Laplacian edge technique of \citet{vandokkum01} and sky emission lines are subtracted using the \citet{kelson03} method. The orders for each arm are then extracted and rectified in wavelength space using a wavelength solution previously obtained from calibration frames. The 2D orders are merged using the errors as a weight in the overlapping regions. We find that the 1D extraction from the ESO pipeline produces noisy data and choose to perform our own extractions using the "apall' routine within IRAF and interactively optimising the signal window definition and background regions for subtraction. 

Based on our experience, removing the quasar trace matters mostly when the angular separation with the target is less than 2". For this data sample, in the case of J045647.17+040052.9, the angular separation is small (b=0.8") and requires a particular treatment. Similarly, in two cases (Q0826$-$2230 and SDSS J122836.8+101841.7), the quasar counter-image is $\sim$2" away from the targeted absorbing-galaxy. In these cases, we perform a Spectral Point Spread Function (SPSF) fit to remove the trace from the bright quasar. Details of the method are provided in Rahmani et al. (in preparation), but in short, the quasar is fitted with a Moffat function at each wavelength, except in a small window where emission lines from the faint galaxies are expected. In these regions, the profile is interpolated from either edge. We note in addition that this technique allows one to search for faint absorbing-galaxies in exposures where the bright quasar impose too high a contrast. Finally, for each exposure, we extract a quasar spectrum and, when detected, a spectrum of the absorbing-galaxy. We correct the wavelength calibration to a vacuum heliocentric reference. We then merge the 1D spectra weighting each spectrum by the signal-to-noise ratio.

The spectra were flux-calibrated using a spectrophotometric standard star. In some cases, the NIR and UVB spectrum is scaled to the VIS spectrum to match in the overlapping regions. We calculate that the corrections for slit losses are negligible for these observations. We also note that X-Shooter was not designed nor is it operated in purely  photometric conditions. For these reasons, the flux calibration is mostly used to remove instrumental effects.

\section{Analysis}
\subsection{SINFONI Detection}

\begin{table*}
\begin{center}
\caption{Summary of the SINFONI detection and upper limits for the new targets from this work. }
\label{t:results_SINFO}
\begin{tabular}{cccccccc}
\hline\hline
Quasar 		  &\zabs &F(H-$\alpha$){\bf $^b$} &Lum(H-$\alpha$) &SFR  &Id Method 	&References \\
 		    &  &[erg/s/cm$^2]$ &[erg/s]&[M$_{\odot}$/yr]  & &\\
\hline
SDSS J002133.27+004300.9         &0.94187$^a$    & 1.8$\pm 1.1 \times$10$^{-16}$		&8.1$\pm 4.8\times$10$^{41}$    &3.6$\pm$2.2 &proximity	&Rao et al. 2011 \\
Q0052+0041                       &0.7402    &$<$1.6$\times$10$^{-17}$		&$<$0.40$\times$10$^{41}$	    &$<$0.18 	   &spectroscopy	&Lacy et al. 2003\\
SDSS J011615.51-004334.7         &0.9127    &$<$0.7$\times$10$^{-17}$		&$<$0.30$\times$10$^{41}$	    &$<$0.13 	   &photo-z	&Rao et al. 2011\\
J0138-0005                        &0.7821    &$<$1.9$\times$10$^{-17}$		&$<$0.55$\times$10$^{41}$	    &$<$0.24 	   &proximity	&Rao et al. 2011\\
J045647.17+040052.9		 &0.8596    &$<$1.9$\times$10$^{-17}$		&$<$0.70$\times$10$^{41}$	    &$<$0.31 	   &proximity	&Le Brun et al. 1997\\
Q0826-2230                       &0.9110    &$<$0.1$\times$10$^{-17}$		&$<$0.04$\times$10$^{41}$	    &$<$0.02 	   &proximity	&Meiring et al. 2011\\
SDSS J152102.00-000903.1         &0.9590    &$<$2.1$\times$10$^{-17}$		&$<$1.01$\times$10$^{41}$	    &$<$0.45 	   &proximity	&Rao et al. 2011\\
\hline\hline 				       			 	 
\end{tabular}			       			 	 
\end{center}			       			 	 
\vspace{0.2cm}
\begin{minipage}{180mm}
{\bf Note:} \\
{\bf $^a$:} In the case of SDSS J002133.27+004300.9, \zabs is derived from the \ha\ emission of the absorbing-galaxy observed with SINFONI.\\
{\bf $^b$:} The 2.5-$\sigma$ upper limits for non-detections are computed for an unresolved source spread over 32 spatial pixels and spectral FWHM = 6 pixels = 9 \AA. \\
\end{minipage}
\end{table*}

\begin{figure}
\begin{center}
\includegraphics[height=6.cm, width=8.cm, angle=0]{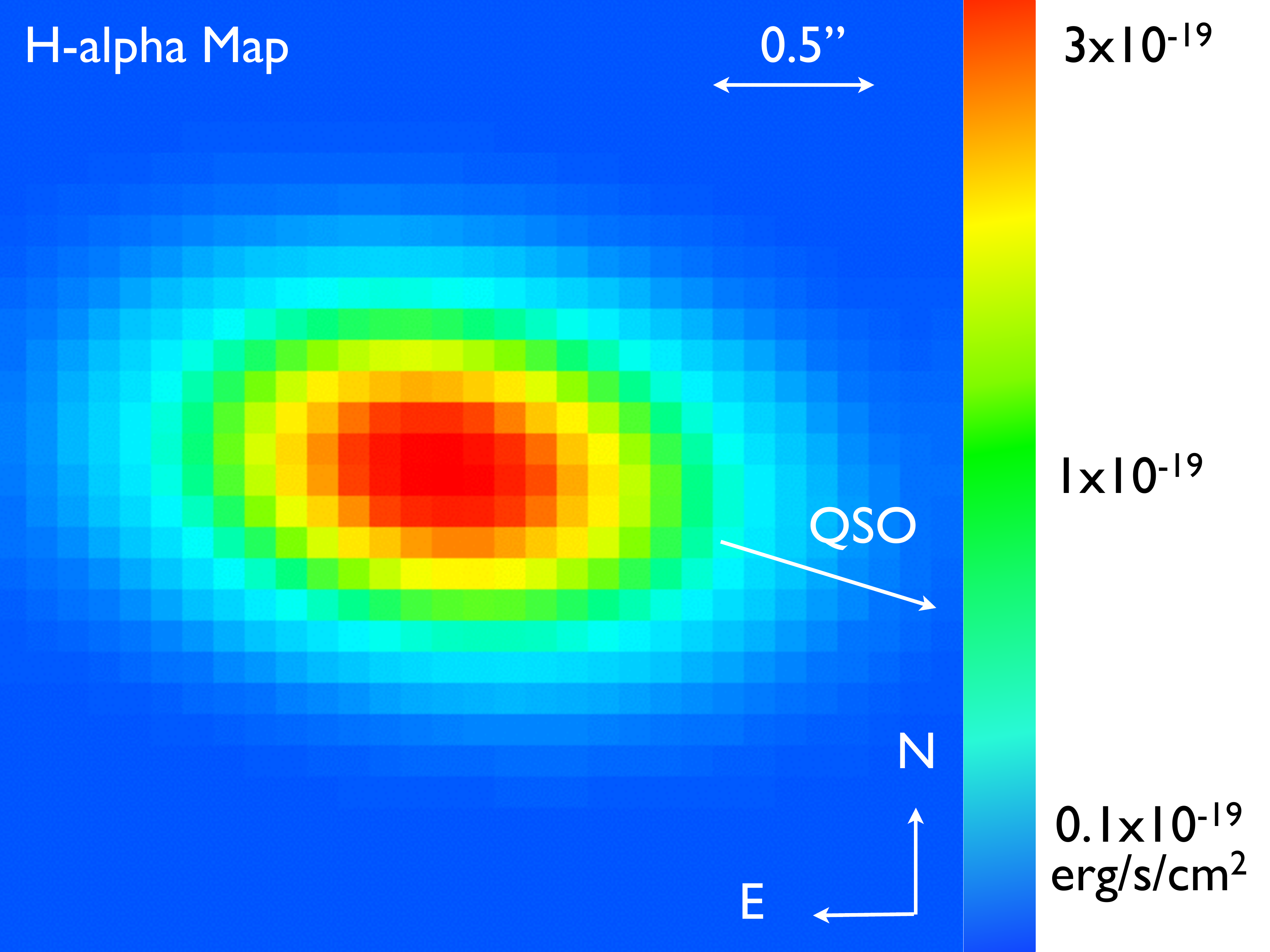}
\caption{{\bf \ha\ flux map of the sub-DLA-galaxy towards SDSS J002133.27+004300.9 at  \zabs=0.9420 convolved with the model.} The quasar is not covered by our data. The orientation and scales are indicated on the figures as well as the direction to the quasar. 
}
\label{f:Q0021_HaMap}
\end{center}
\end{figure}

\begin{figure}
\begin{center}
\includegraphics[height=7.cm, width=5.5cm, angle=-90]{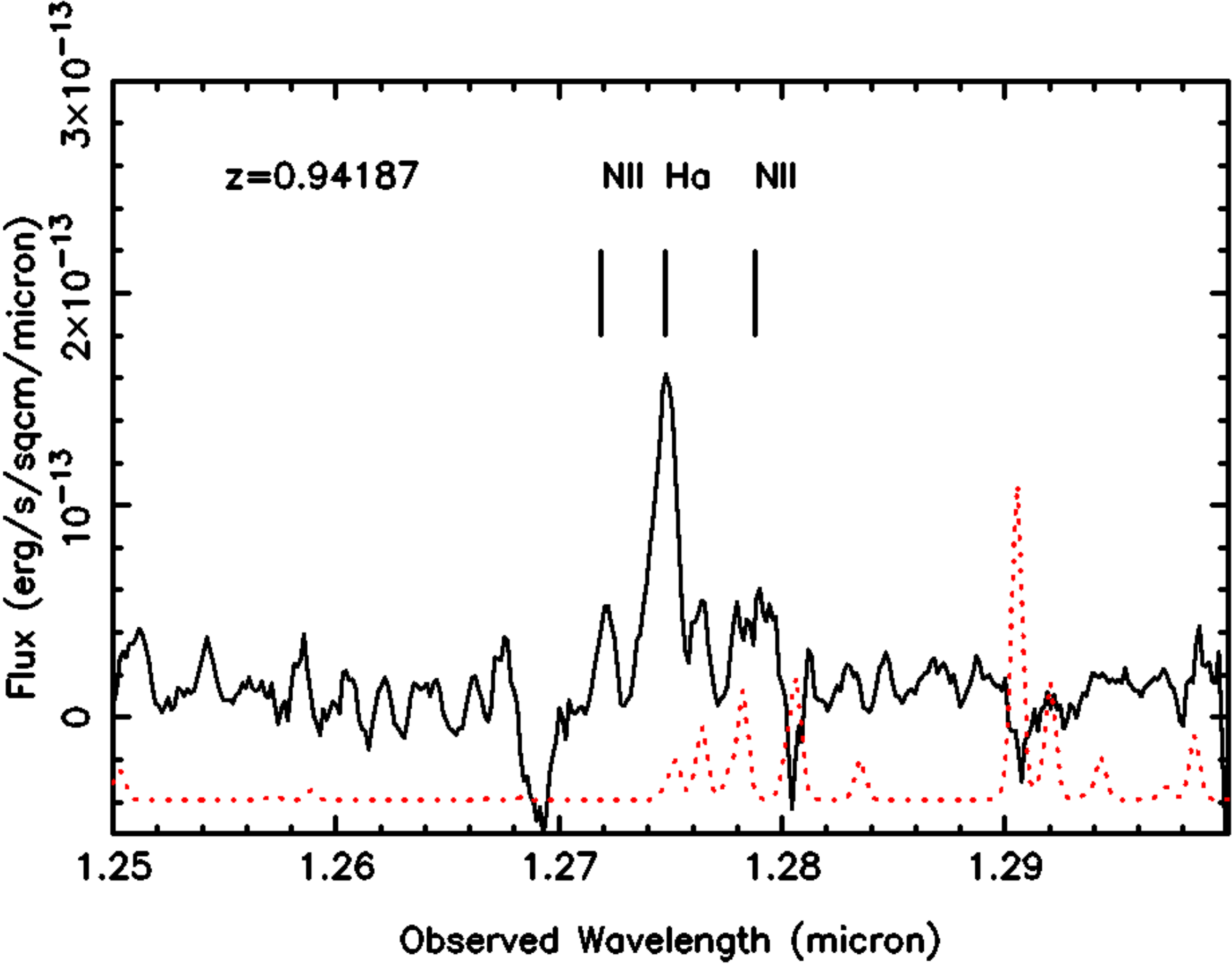}
\caption{{\bf 1D SINFONI \ha\ line detection of the absorbing-galaxy toward SDSS J002133.27+004300.9.} The integrated spectrum  over 13 pixels circular aperture of the redshifted H$\alpha$ emission line of the absorbing-galaxy. The units are in erg/s/cm$^2$/$\mu m$. The spectrum is smoothed (5 pixel boxcar). The dotted spectrum at the bottom of the panel is the sky spectrum with arbitrary flux units, scaled for clarity, and indicating the position of the OH sky lines.  
}
\label{f:Q0021_EmLines_SINFO}
\end{center}
\end{figure}

We report just one detection with SINFONI in the line-of-sight towards the quasar SDSS J002133.27+004300.9. The remaining fields do not show H-$\alpha$ at the expected redshifted wavelength even though the galaxies associated with the quasar absorbers were reported in the literature based on photometric or spectroscopic redshift. Note that the SINFONI instrument is not sensitive to continuum detection. References to previous reports in the literature for these absorbing-galaxies are provided in the last column of Table~\ref{t:results_SINFO}. The low detection rate is unexpected from a sample selected on the basis of apriori galaxy identification given that spatial coincidences are too rare unless there is some physical connection. We emphasise that these non-detections are genuine, probably indicating a lower SFR in these objects than our detection limits and possibly high dust attenuation. The limits on the star formation rate (SFR) from the H-$\alpha$ non-detection are also provided in Table~\ref{t:results_SINFO}, overall reaching a threshold below $<$1 M$_{\odot}$/yr. 

In the case of the quasar SDSS J002133.27+004300.9 at z$_{\rm quasar}$=1.245, \citet{rao06} have reported two sub-DLAs. One system at \zabs=0.5023 has a column density of  \lognhi=19.54$^{+0.02}_{-0.03}$ and another at higher redshift \zabs=0.9420 with \lognhi=19.38$^{+0.10}_{-0.15}$. Furthermore, \citet{rao11} have obtained J, H \& K band ground-based imaging of 10$\times$10 arcsec of this field. After performing a careful PSF subtraction of the quasar, these authors detect no object at small impact parameters. However, three objects are detected further out, two of which are seen in all three bands. Based on both these IR images and SDSS photometry, the so-called "object 1" was estimated to be at \zphot=0.549 with an E(B$-$V)=0.50 and a metallicity Z=0.004, consistent with the low-redshift sub-DLA listed in Table~\ref{t:JoO_XSH}. For the "object 2", no detections are made in SDSS, but the red colours point to a late-type galaxy at z$\sim$1, therefore more likely matching the higher-redshift sub-DLA. Indeed, our SINFONI observations confirm that this galaxy's redshift corresponds to the one from the \zabs=0.9420 sub-DLA. We report the detection of the \ha\ emission line at z=0.94187. Figure~\ref{f:Q0021_HaMap} shows the H$\alpha$ flux map of this galaxy convolved with the model. We measure an H-$\alpha$ flux of 1.8$\pm 1.1\times$10$^{-16}$ erg/s/cm$^2$, corresponding to a luminosity L(H-$\alpha$)=8.1$\pm 4.8 \times$10$^{41}$ erg/s (see Table~\ref{t:results_SINFO}). We then derive the SFR assuming the \citet{kennicutt98} flux conversion corrected to a Chabrier (2003) IMF, leading to SFR=3.6$\pm$2.2 M$_{\odot}$/yr (no dust correction) at an angular separation of 10.8" (corresponding to an impact parameter of 85 kpc). Figure~\ref{f:Q0021_EmLines_SINFO} shows the 1D integrated spectrum extracted from the SINFONI cube and zoomed around the \ha\ emission line. While \ha\ is clearly detected, \nii\ is not seen with a flux limit less than 3.97$\times$10$^{-16}$ erg/s/cm$^2$. Using the N2-parameter \citep{pettini04} based on our limit on the \nii $\lambda$ 6585/H-$\alpha$ ratio, we can derive a limit on the emission metallicity: 12+log(O/H)$<$9.10, which is not very constraining in this case.

To estimate the kinematics of the host galaxy,
we use the GalPak$^{\rm 3D}$ algorithm \citep{bouche15}
which compares directly the datacube with a parametric model mapped in
$x,y,\lambda$ coordinates. The algorithm uses a Markov Chain Monte
Carlo (MCMC) approach with a non-traditional proposed distribution in order to efficiently probe
the parameter space.
This algorithm is able to recover  the morphological parameters
(inclination, position angle) to within 10\%\ and the kinematic
parameters (maximum rotation velocity) within 20\%, irrespective of
the seeing (up to 1.2") provided that the maximum signal-to-noise
(SNR)  is greater than $\sim3$ pix$^{-1}$ and that  the galaxy
half-light radius to seeing radius   is greater than about 1.5.
In our case, the quality of the data for that field is limited due to the fact that only one OB was executed (and classified as "terminated" by ESO due to degrading seeing to 0.95" during exposure), resulting with a seeing of 0.75". As a result, the SNR in the brightest pixel is about 2.7, i.e.
just below the threshold of 3, and a careful look at the MCMC chain is
required.

Using a Gaussian flux profile and an arctan velocity profile,
we find that its dispersion $\Sigma_o$ is well constrained at around
$\sigma$=123$\pm$11 km/s, while the maximum circular velocity is essentially
poorly constrained (V$_{\rm max}$=80$\pm$40 km/s). The half-light radius has converged to
about 6 kpc,
and the PA is found to be 60-90deg, while the
inclination and turn-over-radius are both unconstrained.

Assuming the system towards SDSS J002133.27+004300.9 is rotating, we can use the enclosed mass to determine the dynamical mass within r$_{1/2}$ (Epinat et al. 2009):

\begin{equation}
M_{\rm dyn} = V_{max}^2~r_{1/2}~/~G
\end{equation}

where $V_{max}$ and r$_{1/2}$ are measured from the 3D kinematical fit to the SINFONI cube. We therefore find M$_{dyn}$=10$^{9.9\pm0.4} $M$_{\odot}$.

We are able to estimate the mass of the halo in which the system towards SDSS J002133.27+004300.9 resides, assuming a spherical virialised collapse model (Mo \& White 2002):

\begin{equation}
M_{\rm halo}= 0.1 H_o^{-1} G^{-1} \Omega_m^{-0.5} (1+z)^{-1.5} V_{max}^3
\end{equation}

using inclination-corrected V$_{max}$ value. We find M$_{halo}$=10$^{11.9\pm0.5}$ M$_{\odot}$. This halo mass is comparable with the one from the Milky Way: 1.9$^{+3.6}_{-1.7} \times$ 10$^{12}$ M$_{\odot}$ (Wilkinson et al. 1999).

\subsection{X-Shooter Detections}

\begin{table*}
\begin{center}
\caption{Summary of the X-Shooter detections from the new targets in this work.}
\label{t:results_XSH}
\begin{tabular}{ccccccc}
\hline\hline
Quasar 		  &\zabs &Detected? &Flux at \ha\ position$^b$ &Id Method   	&References \\
 		    &  & &[erg/s/cm$^2]$ & &\\
\hline
SDSS J002133.27+004300.9         &0.94187$^a$    &detected			&1.76$\times$10$^{-16}$	    &proximity    &Rao et al. 2011 \\
Q0052+0041                       &0.7402    	 &bright continuum		&2.0$\times$10$^{-18}$	    &spectroscopy &Lacy et al. 2003\\
SDSS J011615.51-004334.7         &0.9127    	 &faint continuum		&1.2$\times$10$^{-18}$	    &photo-z 	  &Rao et al. 2011\\
J045647.17+040052.9		 &0.8596    	 &undetected (b=0.8")		&--			    &proximity 	  &Le Brun et al. 1997\\
Q0826-2230                       &0.9110   	 &undetected 			&--	    		    &proximity 	  &Meiring et al. 2011\\
SDSS J122836.8+101841.7          &0.9376    	 &faint continuum		&1.2$\times$10$^{-18}$	    &photo-z 	  &Rao et al. 2011\\
SDSS J152102.00-000903.1         &0.9590    	 &very faint continuum		&0.5$\times$10$^{-18}$	    &proximity 	  &Rao et al. 2011\\
\hline\hline 				       			 	 
\end{tabular}			       			 	 
\end{center}			       			 	 
\vspace{0.2cm}
\begin{minipage}{180mm}
{\bf Note:} \\
{\bf $^a$:} In the case of SDSS J002133.27+004300.9, \zabs\ is derived from the \ha\ emission of the absorbing-galaxy observed with SINFONI.\\
{\bf $^b$:} The values are fluxes in the continuum detected with X-Shooter averaged over a dozen pixels, except for SDSS J002133.27+004300.9 which is the detected flux in the detected emission line. These fluxes are significantly below the SINFONI detection limits (see Table~\ref{t:results_SINFO}).\\
\end{minipage}
\end{table*}			       			 	 

\begin{figure}
\begin{center}
\includegraphics[height=6.5cm, width=9.5cm, angle=0]{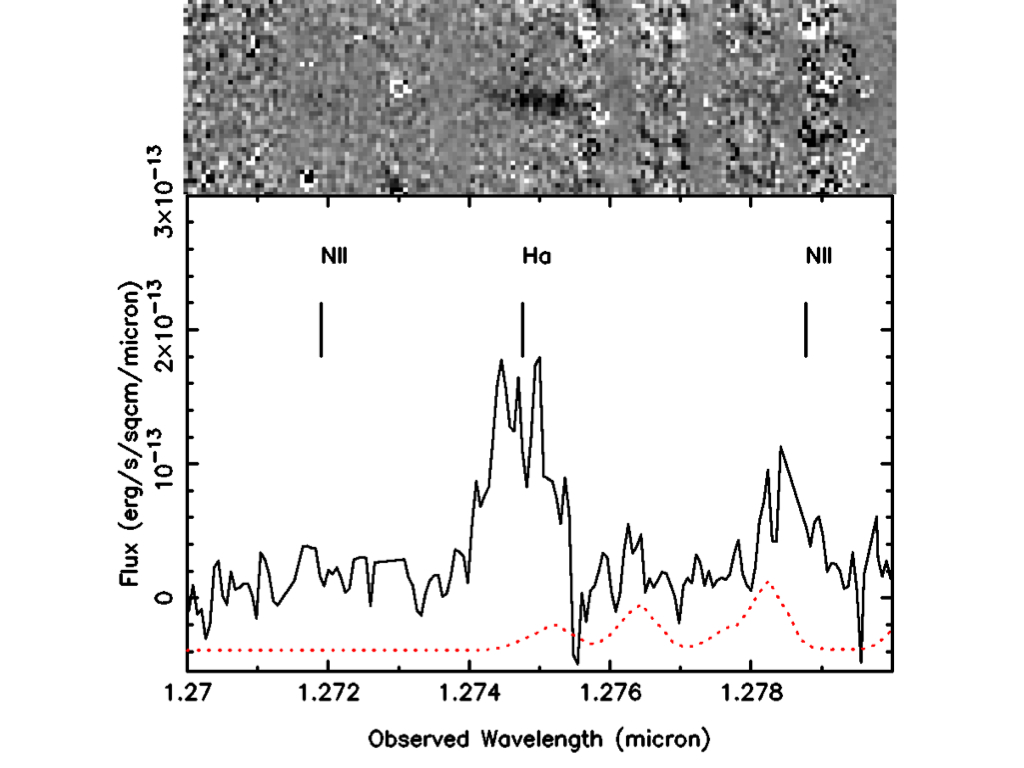}
\caption{{\bf 1D X-Shooter NIR arm revealing the \ha\ line in the absorbing-galaxy toward SDSS J002133.27+004300.9.} The units are in erg/s/cm$^2$/$\mu m$. The spectrum is not smoothed. The dotted spectrum at the bottom of the panel is the sky spectrum with arbitrary flux units, scaled for clarity, and indicating the position of the OH sky lines. The top panel shows the 2D emission line on the same wavelength scale. 
}
\label{f:Q0021_EmLines_XSH}
\end{center}
\end{figure}

\begin{figure}
\begin{center}
\includegraphics[height=8.5cm, width=6.5cm, angle=-90]{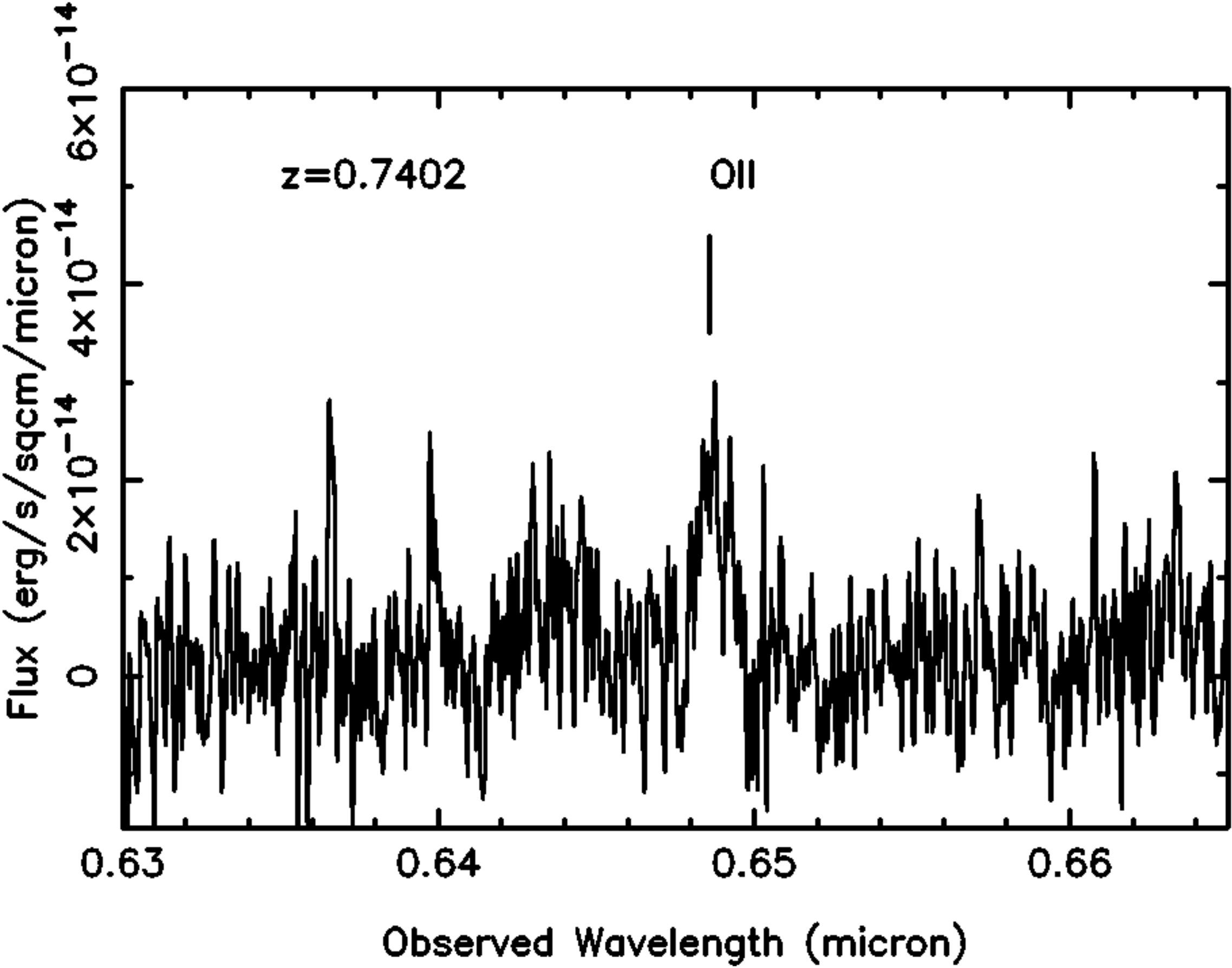}
\caption{{\bf 1D X-Shooter VIS arm revealing \oii\ $\lambda$ 3727 line in the absorbing-galaxy toward Q0052+0041.} The units are in erg/s/cm$^2$/$\mu m$. The spectrum is smoothed (5 pixel boxcar). From a careful inspection of the individual exposures, we can tell that the other narrower features bluer of \oii\ are spurious. 
}
\label{f:Q0051_EmLines_XSH}
\end{center}
\end{figure}

Our X-Shooter observations include several detections of the absorbing-galaxies. We detail our findings for each individual objects in turn below. Table~\ref{t:results_XSH} summarises the results.

 \vspace{0.3cm}
{\bf SDSS J002133.27+004300.9:} In the section above, we reported detection of \ha\ at the expected position of the absorbing-galaxy in this field. The X-Shooter observations confirm this detection. The \ha\ line is clearly seen in each 2D frame from individual OBs. The detections are even stronger in the most recent observations (taken in July 2015) notwithstanding that the requested constraints on the observing conditions are similar. An excerpt 2D frame is shown in the top panel of Fig~\ref{f:Q0021_EmLines_XSH}. The feature is clearly extended along the dispersion axis. The bottom panel of Fig~\ref{f:Q0021_EmLines_XSH} shows the 1D extracted spectrum together with a sky spectrum (red dotted line). We measure a H-$\alpha$ flux of F(\ha)=1.76$\times$10$^{-16}$ erg/s/cm$^2$, almost identical to our measurement from SINFONI observations (see Table~\ref{t:results_SINFO}). The agreement between the two measurements is remarkable given the uncertainties in flux calibration inherent to the X-Shooter instrument. 

There are strong hints of the presence of the \nii\ $\lambda$ 6585.27 emission line too (see Fig~\ref{f:Q0021_EmLines_XSH}). However, we notice the presence of an OH sky line at this wavelength. For this reason, we conservatively report a limit on the \nii\ flux of F(\nii)$<$1.7$\times$10$^{-18}$ erg/s/cm$^2$. Similarly, we report $<$1.1$\times$10$^{-18}$ erg/s/cm$^2$ from the non-detection of the \nii\ $\lambda$ 6549.86 in the X-Shooter NIR arm. In the VIS arm, the non-detection of both \oiii\ and \oii\ doublets results in the following flux limits: F(\oiii\ $\lambda$ 5008)$<$0.7$\times$10$^{-18}$, F(\oiii\ $\lambda$ 4960)$<$0.7$\times$10$^{-18}$, F(\oii\ $\lambda$ 7243)$<$0.8$\times$10$^{-18}$ and F(\oii\ $\lambda$ 7237)$<$0.8$\times$10$^{-18}$ erg/s/cm$^2$.
 
 \vspace{0.3cm}
{\bf  Q0052+0041:} \citet{lacey03} used a HST spectrum of this quasar to measure a \hi\ column density of \lognhi=20.4$\pm$0.1. In addition, \citet{lacey03} reported the presence of the absorbing-galaxy in this field based on both NIR imaging and a Keck/LRIS spectrum showing \oii\ emission and Ca H\&K absorption. The authors report that the object is a fairly normal, star-forming galaxy with a luminosity of approximately L$_*$.  The \oii\ and Ca H\&K lines present a slight velocity offset which is also reflected in the absorption profiles of \mgii, \mgi\ and \feii. It is noted that these two components might either correspond to two distinct galaxies, one of which is a dwarf which remains undetected. Alternatively, this shift could be due to motion of internal gas in the main galaxy. \citet{lacey03} do not measure the flux in the \oii\ line or give an estimate for the SFR, however.  

Based on the X-Shooter observations, we report the detection of a bright continuum at the expected position of the absorbing-galaxy. As in our SINFONI observations, we do not detect \ha\ in that object in NIR arm of X-Shooter (down to F(\ha)$<$1.6$\times$10$^{-17}$ erg/s/cm$^2$). However, we do report \oii\ at the expected position of the absorbing-galaxy based on the VIS arm of the X-Shooter spectrum. The \oii\ doublet is unresolved
notwithstanding the X-Shooter spectral resolution. The Ca H\&K are not detected, but we note that the SNR of our spectrum is lower than the Keck/ESI spectrum of Lacy et al. (2003). Fig~\ref{f:Q0051_EmLines_XSH} shows the 1D extracted spectrum showing a rather broad emission. Based on this detection, we derive F(\oii)$=$1.9$\pm 0.3 \times$10$^{-17}$erg/s/cm$^2$, which translates into a luminosity of L(\oii)$=$3.8 $\pm$ 0.6$\times$10$^{41}$erg/s. Using the prescription from \citet{kennicutt98} :

\begin{equation}
SFR_{[O II]} = 1.4 \times 10^{41} L([O II])
\end{equation}

we derive a SFR$=$5.3$\pm$0.7 M$_{\odot}$/yr uncorrected for dust depletion. This estimate is well above the limit derived with SINFONI based on the non-detection of \ha\ (SFR$<$0.18 M$_{\odot}$/yr). However, we note that \cite{moustakas06} report a large scatter in the SFR/\oii\ relation. In fact, the \oii/\ha\ ratio we derive from the non-detection of \ha\ in this system is extreme and falls away from the relation, which might explain the difference in SFR estimates and which leads to questions about the use of the relation. 

 \vspace{0.3cm}
{\bf SDSS J011615.51-004334.7:} \citet{rao06} report a sub-DLA at \zabs=0.9127 along this line-of-sight. They measure \lognhi=19.95$^{+0.05}_{-0.11}$. In addition, \citet{rao11} used J, H and K ground-based images with Sloan colours of the brightest object detected in the field to perform a photometric redshift estimate. The authors note that their estimate is not in agreement with the photometric redshift in the Sloan database but that it does marginally agree with the absorption redshift. 

In the NIR arm of our X-Shooter observation, we report a faint continuum at the expected position of the absorbing-galaxy candidate. The object presents a power-law continuum and an emission line consistent with the redshift of the background quasar (\zem=1.275), but this could be due in part to contamination from the nearby quasar although it is 8.1" away. The flux at the \ha\ position averaged over a dozen pixels is F(\ha)$<$1.2$\times$10$^{-18}$ erg/s/cm$^2$, consistent with the non-detection in our SINFONI observations.

 \vspace{0.3cm}
{\bf J045647.17+040052.9:} There are two absorbers reported along this line-of-sight: a \mgii\ and \civ\ absorber at \zabs=1.1536 \citep{steidel92} and a DLA with \lognhi=20.75$\pm$0.03 at \zabs=0.8596 \citep{steidel95}. Using ground-based imaging, Steidel et al. (1995) report an absorbing-galaxy candidate at an angular separation of 2.1". However, \citet{lebrun97} used HST images and report a faint galaxy at an angular separation of 0.8". \citet{lebrun97} argue that the difference in impact parameters is due to their less accurate quasar PSF subtraction than in the ground-based images of \citet{steidel95}. 

We target the object at angular separation of 0.8" with our X-Shooter observations but do not report any detections. We note that the small angular separation (b=0.8") might have hampered detection in this case. On top of that, the quasar is significantly brighter than in the other fields (V mag = 16.5), which might limit our capability to detect an object several order of magnitudes fainter. In fact, we find a weak detection at the expected position of \ha\ line but the \ha\ flux is less than a percent of quasar flux, so we do not claim a detection. We note that \citet{takamiya12} put a 3$-\sigma$ upper limit from non-detection of the \oii\ emission line flux for the z=0.8596 DLA of F(\oii) $< 8.3 \times 10^{-19}$ erg s$^{-1}$ cm$^{-2}$ assuming a line width of 100 km s$^{-1}$. They also report an emission-line galaxy at z=0.0715 with multiple emission lines and a new \mgii\ absorber at z=1.245. 

 \vspace{0.3cm}
{\bf Q0826-2230:} The sub-DLA at \zabs=0.9110 has its \nhi\ measured by \citet{rao06} to be \lognhi=19.04$\pm$0.04. Meiring et al. (2011) used Sloan and SOAR imaging to obtain photometric redshifts of the absorbing-galaxy candidates in the field. The images show that the quasar is lensed. The closest object, "object 1" is identified as a star. The next two closest galaxies ("objects 4 and 5") have photometric redshifts which are not consistent with the redshift of the absorber. In our X-Shooter observations we target "object 4". In this case too, we do not report any detections in our X-Shooter observations. Again, as in the case of J045647.17+040052.9, we note that the quasar is significantly brighter than in the other fields (V mag = 16.2) creating a high contrast.  Recently, Straka et al. (submitted) determine spectroscopically that "objects 4 and 5" are indeed {\it not} at the redshift of the absorber.

\vspace{0.3cm}
{\bf SDSS J122836.8+101841.7:} \citet{rao06} report a sub-DLA at \zabs=0.9376 along this line of sight. They measure \lognhi=19.41$^{+0.12}_{-0.18}$. In addition, \citet{rao11} observed J, H and K ground-based images. They detect two objects in the field and find the photometric redshift of the closer candidate to be consistent with the absorber.  

In the NIR arm of our X-Shooter observation, we report a faint continuum at the expected position of the absorbing-galaxy candidate. While this continuum is clearly visible in the 2D images, a careful visual inspection does not show predominant emission lines, rendering its identification and redshift determination difficult. To compare with SINFONI observations, we measure the flux averaged over a dozen pixels in the continuum at the expected position of the \ha\ emission line (F(\ha)$<$1.2$\times$10$^{-18}$ erg/s/cm$^2$). This flux is significantly below the SINFONI detection limit (see Table~\ref{t:results_SINFO}). In case this object were an early type and bulge-dominated galaxy, we would not expect to detect its \ha\ emission, but only its continuum. This hypothesis could be checked with further imaging or deeper spectroscopy. Indeed, Rahmani et al. (in prep) report two such examples where the redshift of the absorbing-galaxy is confirmed from absorption lines in the galaxy spectra. 

\vspace{0.3cm}
{\bf SDSS J152102.00-000903.1:} \citet{rao06} report a sub-DLA at \zabs=0.9590 along this line of sight. They measure \lognhi=19.40$^{+0.08}_{-0.14}$. In addition, \citet{rao11} observed J, H and K ground-based images. They detect two objects in the field, but could not derive photometric redshifts for either. They note the object closer to the quasar has colours consistent with being an early type galaxy at the absorption redshift and so we targeted that object with our X-Shooter observations. 

In the NIR arm, we report a very faint continuum at the expected position of the absorbing-galaxy candidate. The flux at the \ha\ position averaged over a dozen pixels is F(\ha)$<$0.5$\times$10$^{-18}$ erg/s/cm$^2$, consistent with the non-detection in our SINFONI observations.

\subsection{Quasar Absorption Spectroscopy}

\begin{table}
\begin{center}
\caption{Voigt profile fit parameters to the low- and intermediate- ionisation species for the z=0.94187 log N(H\,I)=$19.38^{+0.10}_{-0.15}$ absorber towards SDSS J002133.27+004300.9.}
\label{t:fit_col}
\begin{tabular}{ccccc}
\hline
\hline
Comp. & $z_{abs}$ & b & Ion & log $N$ \\
 & & km $s^{-1}$ & & cm$^{-2}$ \\
 \hline
1 & $0.94072$ & $8.2\pm0.4$ & FeII & $12.47\pm0.07$\\
   &   &   & ZnII & $-$\\
   &   &   & CrII & $11.82\pm1.13$\\
   &   &   & SiII & $-$\\
   &   &   & MgI & $10.53\pm0.52$\\
   &   &   & AlIII & $12.21\pm0.14$\\
   &   &   & MnII & $-$\\
   &   &   & MgII & $13.09\pm0.02$\\
2 & $0.94087$ & $5.6\pm0.5$ & FeII & $11.80\pm0.013$\\
   &   &   & ZnII & $12.26\pm0.06$\\
   &   &   & CrII & $-$\\
   &   &   & SiII & $-$\\
   &   &   & MgI & $-$\\
   &   &   & AlIII & $11.53\pm0.55$\\
   &   &   & MnII & $-$\\
   &   &   & MgII & $12.66\pm0.02$\\
3 & $0.94165$ & $9.0\pm0.2$ & FeII & $14.08\pm0.03$\\
   &   &   & ZnII & $-$\\
   &   &   & CrII & $12.42\pm0.18$\\
   &   &   & SiII & $14.99\pm0.11$\\
   &   &   & MgI & $12.27\pm0.01$\\
   &   &   & AlIII & $13.88\pm0.08$\\
   &   &   & MnII & $11.92\pm0.16$\\
   &   &   & MgII & $-$\\
4 & $0.94183$ & $2.8\pm1.0$ & FeII & $12.53\pm0.05$\\
   &   &   & ZnII & $11.67\pm0.11$\\
   &   &   & CrII & $-$\\
   &   &   & SiII & $-$\\
   &   &   & MgI & $-$\\
   &   &   & AlIII & $12.17\pm0.15$\\
   &   &   & MnII & $-$\\
   &   &   & MgII & $-$\\
5 & $0.94192$ & $2.5\pm0.5$ & FeII & $12.83\pm0.07$\\
   &   &   & ZnII & $11.88\pm0.07$\\
   &   &   & CrII & $12.18\pm0.25$\\
   &   &   & SiII & $-$\\
   &   &   & MgI & $10.80\pm0.15$\\
   &   &   & AlIII & $11.97\pm0.23$\\
   &   &   & MnII & $-$\\
   &   &   & MgII & $-$\\
6 & $0.94200$ & $2.1\pm0.3$ & FeII & $12.94\pm0.08$\\
   &   &   & ZnII & $12.16\pm0.06$\\
   &   &   & CrII & $11.78\pm0.61$\\
   &   &   & SiII & $-$\\
   &   &   & MgI & $10.18\pm0.64$\\
   &   &   & AlIII & $11.47\pm0.56$\\
   &   &   & MnII & $-$\\
   &   &   & MgII & $-$\\
7 & $0.94208$ & $1.9\pm0.6$ & FeII & $12.69\pm0.11$\\
   &   &   & ZnII & $10.59\pm1.16$\\
   &   &   & CrII & $12.07\pm0.33$\\
   &   &   & SiII & $13.64\pm1.16$\\
   &   &   & MgI & $10.68\pm0.19$\\
   &   &   & AlIII & $12.12\pm0.22$\\
   &   &   & MnII & $-$\\
   &   &   & MgII & $-$\\
  \hline\hline
\end{tabular}
\end{center}        
\end{table} 

\begin{table}
\begin{center}
\begin{tabular}{ccccc}
\hline
\hline
Comp. & $z_{abs}$ & b & Ion & log $N$ \\
 & & km $s^{-1}$ & & cm$^{-2}$ \\
 \hline
8 & $0.94214$ & $2.7\pm1.9$ & FeII & $12.39\pm0.10$\\
   &   &   & ZnII & $-$\\
   &   &   & CrII & $-$\\
   &   &   & SiII & $13.80\pm0.83$\\
   &   &   & MgI & $11.11\pm0.09$\\
   &   &   & AlIII & $12.08\pm0.20$\\
   &   &   & MnII & $-$\\
   &   &   & MgII & $-$\\
9 & $0.94233$ & $13.9\pm0.3$ & FeII & $14.41\pm0.04$\\
   &   &   & ZnII & $-$\\
   &   &   & CrII & $12.30\pm0.31$\\
   &   &   & SiII & $14.98\pm0.11$\\
   &   &   & MgI & $12.80\pm0.01$\\
   &   &   & AlIII & $13.62\pm0.04$\\
   &   &   & MnII & $12.20\pm0.10$\\
   &   &   & MgII & $-$\\
   \hline\hline
\end{tabular}
\end{center}        
\end{table} 

\begin{table}
\begin{center}
\caption{{\bf Neutral gas-phase abundances of SDSS J002133.27+004300.9 from UVES quasar spectroscopy}.  These metallicities with respect to solar values are measured in absorption along the line-of-sight to the background quasar. }
\label{t:fit_ab}
\begin{tabular}{ccc}
\hline\hline
Element
        &Column density& Abundance \\
        &of II ions &  \\
\hline
Fe & $14.61\pm0.03$ & $-0.27\pm0.18$ \\
Zn  & $<12.66$& $<+0.57$ \\
Cr  & $<12.93$ & $<-0.09$ \\
Si  & $15.31\pm0.08$ & $+0.42\pm0.23$ \\
Mg  & $13.23\pm0.02$ & $-1.75\pm0.17$ \\
Mn  & $12.38\pm0.09$ & $-0.43\pm0.24$ \\
Al  & $14.10\pm0.05$ & $+0.27\pm0.20$ \\
Ti  & $<12.02$ & $<-0.31$ \\
\hline\hline        
\end{tabular}        
\end{center}        
\begin{minipage}{90mm}
\end{minipage}
\end{table}

\begin{figure*}
\begin{center}
\includegraphics[height=12.5cm, width=19.cm, angle=0]{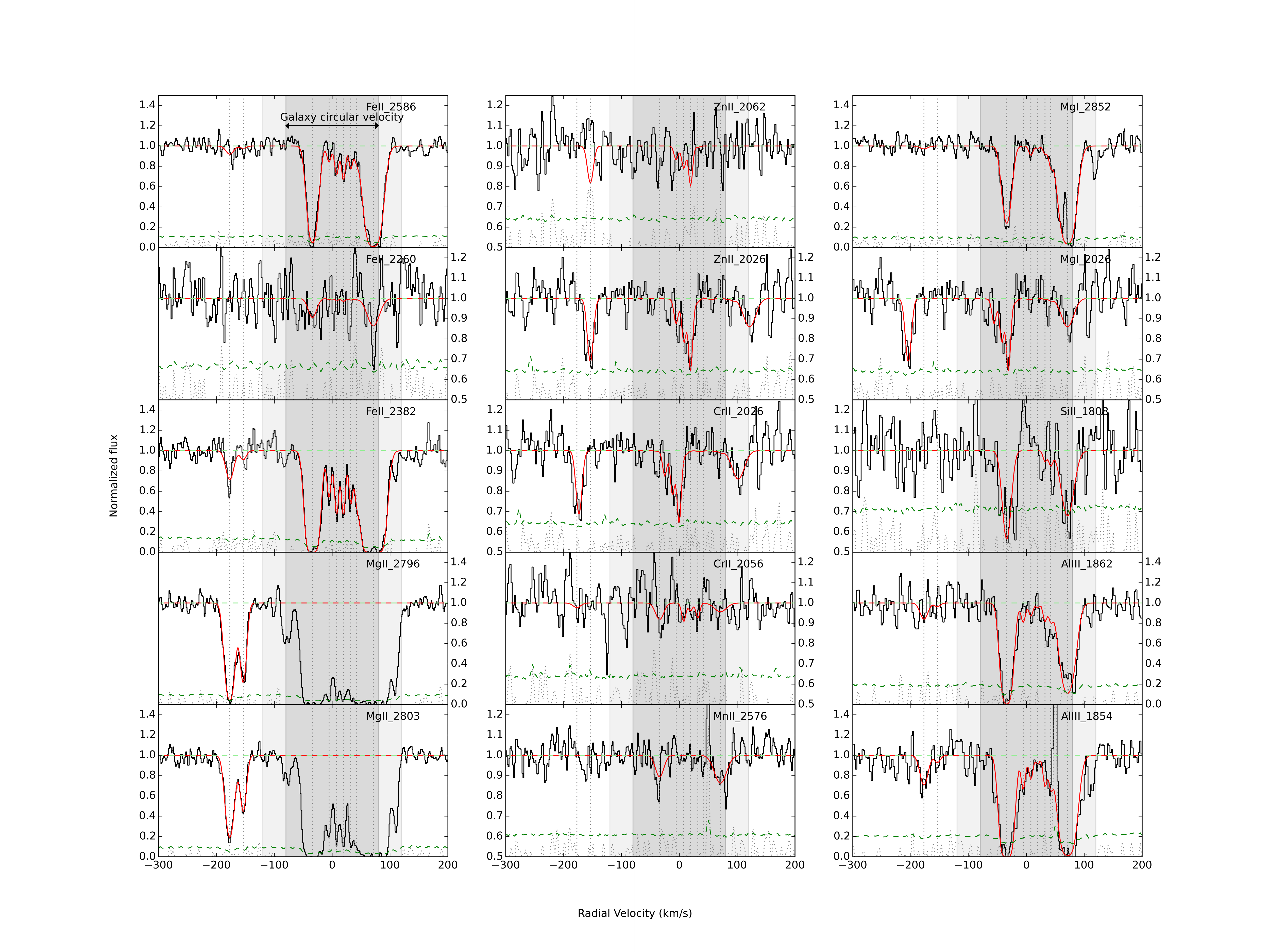}
\caption{{\bf Normalised 1D UVES quasar spectrum of SDSS J002133.27+004300.9.} The normalised UVES spectrum (black) and the Voigt profile fits (red) of the sub-DLA are shown on a velocity scale where 0 km/s is set to be the emission systemic redshift of the detected galaxy, \zabs=0.94187. The positions of the components are indicated by vertical dotted lines. The error array is shown as a dashed green line and the difference between the UVES spectrum and the fit is shown as a dotted light grey line. For weaker lines, the y-axis scale is adjusted and these latter (error + difference) spectra are shifted up by 0.5. The horizontal arrow on the top left panel and the dark and light grey bands in all panels indicate the maximum velocity measured from SINFONI kinematics study, V$_{\rm max}=80\pm40$ km/s. The weakest blue components at $\sim$180 km/s are not explained by the rotation velocity of the disk in that object. }
\label{f:Q0021_UVES}
\end{center}
\end{figure*}

In our sample, just one X-Shooter target (SDSS J002133.27+004300.9) did not include the quasar in the slit due to its large impact parameter. However, this quasar spectrum was observed under programme ESO 078.A-0003 with UVES and the relevant data are part of the Advanced Data Products ESO archive. We use the UVES spectrum observed in Service Mode on November 16, 2006. The object was observed using a combined 390$+$564 nm setting with exposure time lasting 3000 sec. The reduced data are taken from the Phase 3 ESO archive facility \citep{retzlaff14}. The resulting spectra are  corrected to the vacuum heliocentric reference and then combined by weighting each spectrum by signal-to-noise. In order to perform the analysis of the absorption lines associated with the galaxy, the quasar spectrum is normalised using a spline function going through regions devoid of absorption features. 

Voigt profile fits are commonly used to derive the column density of different elements detected in absorption in quasar spectra. In this case, \citet{dessauges09} derive an abundance for iron and an upper limit for zinc: [Fe/H]=$-$0.21$\pm$0.14 and [Zn/H]$<-$0.41 although no details on the fits are provided. In fact, the UVES quasar spectrum of SDSS J002133.27+004300.9 covers \feii\ $\lambda\lambda\lambda\lambda$ 2600 2586 2382 2260, \znii\ $\lambda\lambda$ 2062 2026, \crii\ $\lambda\lambda$ 2056 2026, \mgii\ $\lambda\lambda$ 2796 2803, \mnii\ $\lambda$ 2576,  \mgi\ $\lambda\lambda$ 2852 2026, \aliii\ $\lambda\lambda$ 1862 1854, \siii\ $\lambda$ 1808 and \tiii\ $\lambda$ 3384. Here, we perform a fit of the detected transitions using a Voigt profile fit under the MIDAS/FITLYMAN software. The redshift used is set to be the emission systemic redshift of the detected galaxy, \zabs=0.94187 from the SINFONI observations described above. The different ions share a common absorption profile, and a similar component structure (Doppler parameter and redshift) has been assumed to exist for all species and used throughout the fit. Due to either weak or saturated absorption in most of the covered ions, the components assumed for the fit are chosen based on \feii\ lines for the strongest components and \mgii\ for the weakest components. 

A 9-component fit is used, with two isolated blue components at $\sim-$180 km/s, and two strong components either side of the absorbing-galaxy systemic redshift. The estimate of the \feii\ total column density is consistent with the value reported by \citet{dessauges09} within the errors. The lines of \znii, \crii\ and \mgi\ are fitted together to deal with the blend around $\lambda_{\rm rest}$ = 2026 \AA. We use a two-step approach: the two isolated blue components are fitted separately due to limitations in the number of free parameters in the fitting algorithm. Our fit in the \znii\ and \crii\ region is hampered by the low SNR of the spectrum, so that we only report upper limits: $\log$ N(\znii)$<12.66$, resulting in [Zn/H]$<+0.57$ and $\log$ N(\crii)$<12.93$, resulting in [Cr/H]$<-0.09$. The [Zn/H] value is much less constraining than the one reported by \citet{dessauges09} on the same data set. In fact, we calculate that given the column density of \znii, a SNR$\sim$100 would be required to reach the limit reported by \citet{dessauges09}. We therefore use our [Zn/H] estimate in the following analysis. Finally, \aliii\, \mnii\ and \siii\ are fitted using the 9 components simultaneously. We note that \tiii\ is covered but undetected and derive an upper limit. The doublet of \mgii\ is detected but strongly saturated, so only the weaker components are fitted.

The parameters for the profile fits are provided in Table~\ref{t:fit_col} and fits are shown in Fig~\ref{f:Q0021_UVES}. The horizontal arrow on the top left panel indicates the maximum velocity measured from SINFONI kinematics study, V$_{\rm max}=80\pm40$ km/s. It is interesting to note that the weakest blue components at $\sim$180 km/s are not explained by the rotation velocity of the disk in that object. In addition, we apply the method proposed by \cite{jenkins09} to both the strong red group of components and to the blue group of components following the method described by Quiret et al. (submitted). In short, this method determines a combination of metallicity (from metal and \hi\ column densities) and dust content using the Milky Way sightlines as references for the depletion pattern. Here, we used it to derive the metallicity of group of components for which we do not have a direct measure of \nhi\ due to blending. We find that the metallicity is higher in the red group of components (\nhi+[X/H]=19.5) at the position of the systemic redshift of the absorbing-galaxy than in the blue group of components (\nhi+[X/H]=17.6), consistent with a picture where this latter gas is associated with metal poor accreting gas \citep{bouche13}. 

The total column densities and abundances are with respect to solar values using the convention [X/H]=$\log$(X/H)-$\log$(X/H)$_{\odot}$ and are listed in Table~\ref{t:fit_ab}. Most notably, we derive [Fe/H]=$-0.27\pm0.18$, [Si/H]=$+0.42\pm0.23$ and [Cr/H]$<-0.09$. Our findings indicate that the metallicity of the system is above solar, [Si/H]=+0.42$\pm$0.23, which is common amongst low-redshift, low \nhi\ quasar absorbers \citep[][Quiret et al. submitted]{som15}. We note that in this system, [Al/Fe] is above solar in many components due to anomalously strong \alii\ features. It is not clear whether this is due to photo-ionisation effects, differential depletion, hidden saturation or nucleosynthetic yields differing from the standard picture of uniform yields from Type II SNe and and Type I SNe at a given redshift.

The remaining targets do not have high-resolution (R$\sim$40,000) quasar spectra available. They are however all covered by our X-Shooter observations. In a subsequent work, we will derive detailed abundance estimates for these low-redshift DLAs and sub-DLAs.

\section{Results}
\subsection{Global Properties of the SINFONI Survey}

\begin{figure}
\begin{center}
\includegraphics[height=8.cm, width=8.cm, angle=-90]{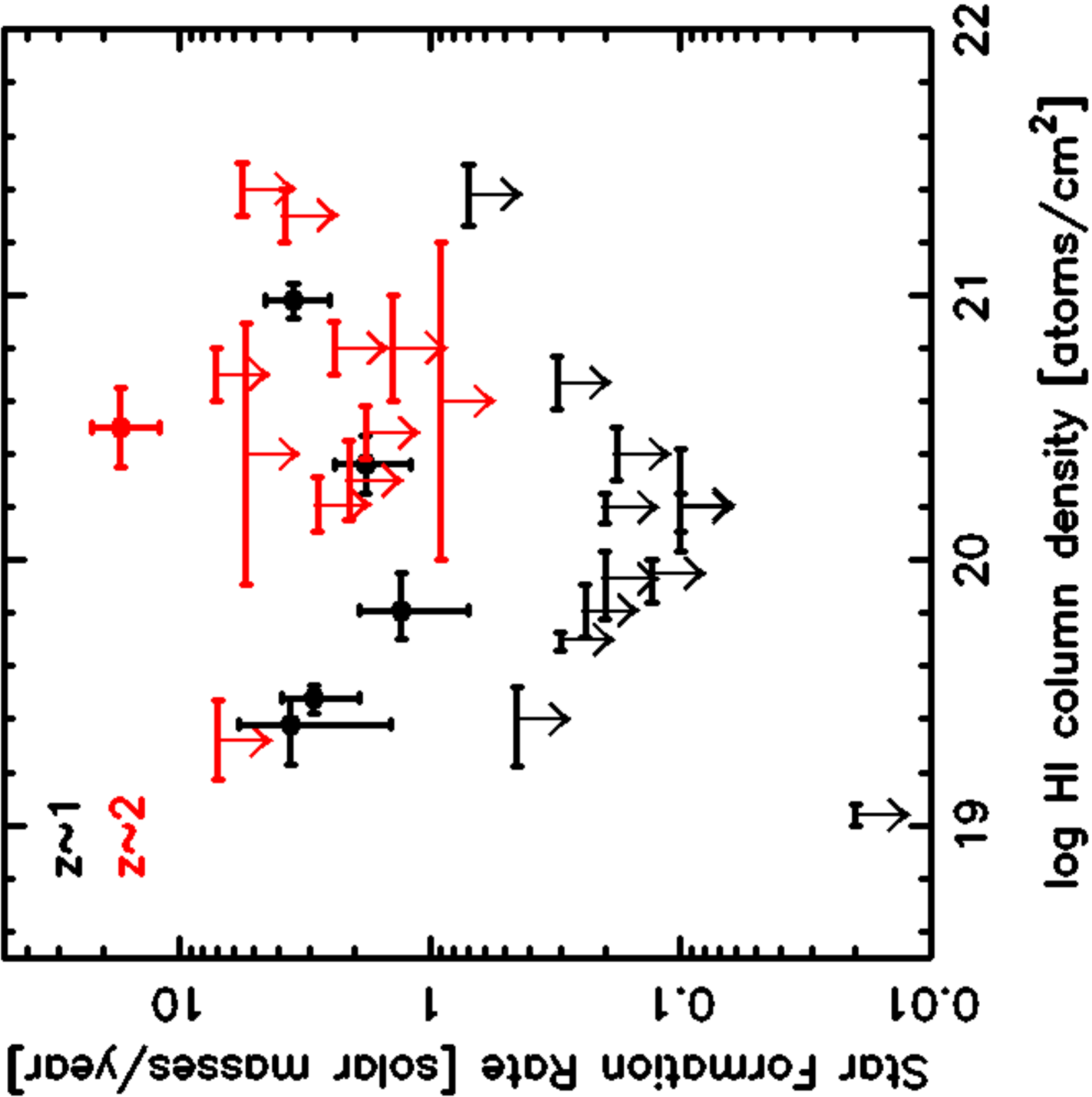}
\caption{{\bf SFR estimates for the full \nhi-selected SINFONI survey.} The black crosses refer to z$\sim$1 targets observed in J-band with SINFONI, while the red crosses are objects lying at z$\sim$2 observed in K-band. }
\label{f:SFR_vs_NHI}
\end{center}
\end{figure}

\begin{figure*}
\begin{center}
\includegraphics[height=5.5cm, width=6cm, angle=-90]{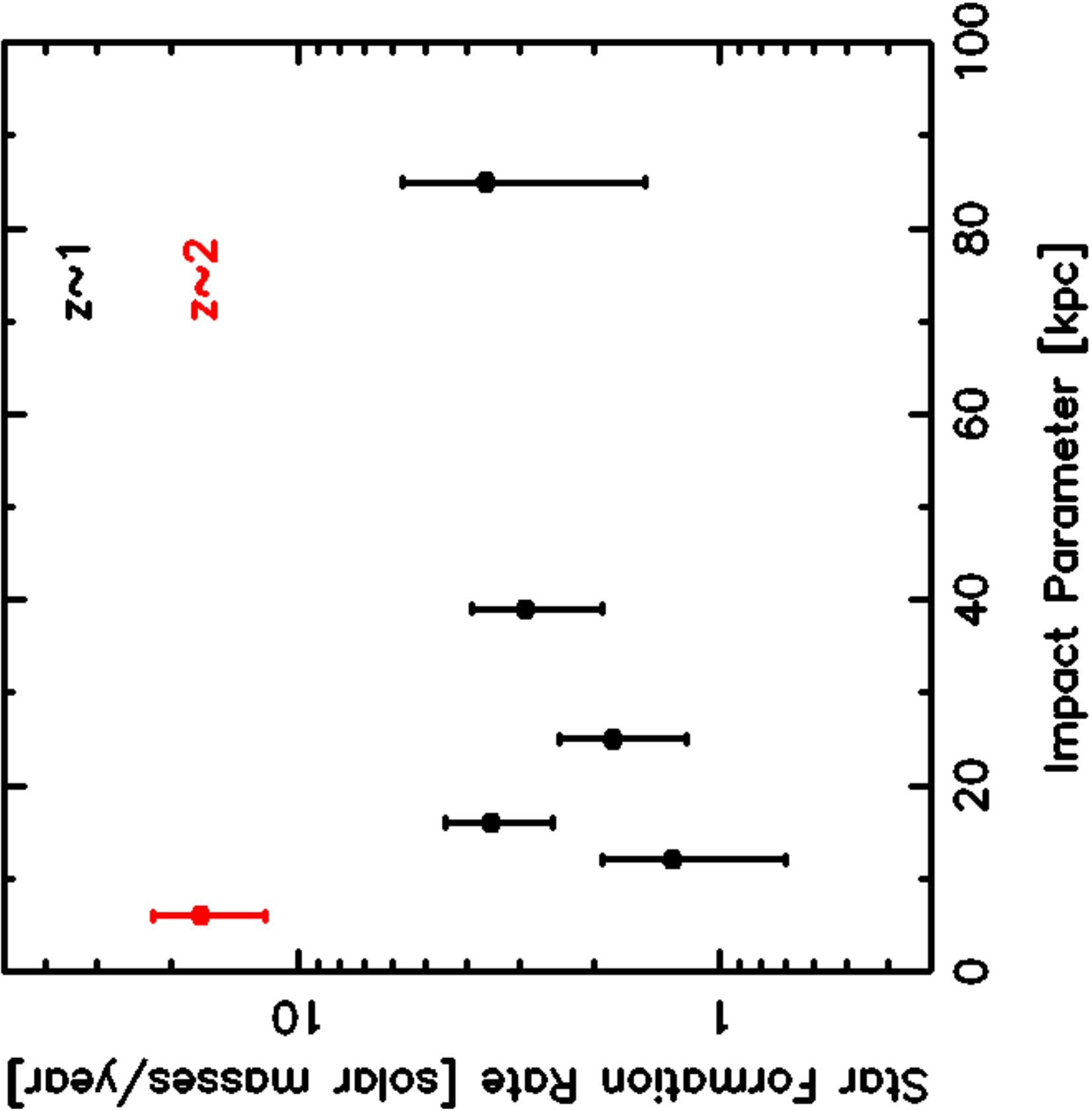}
\includegraphics[height=5.5cm, width=6cm, angle=-90]{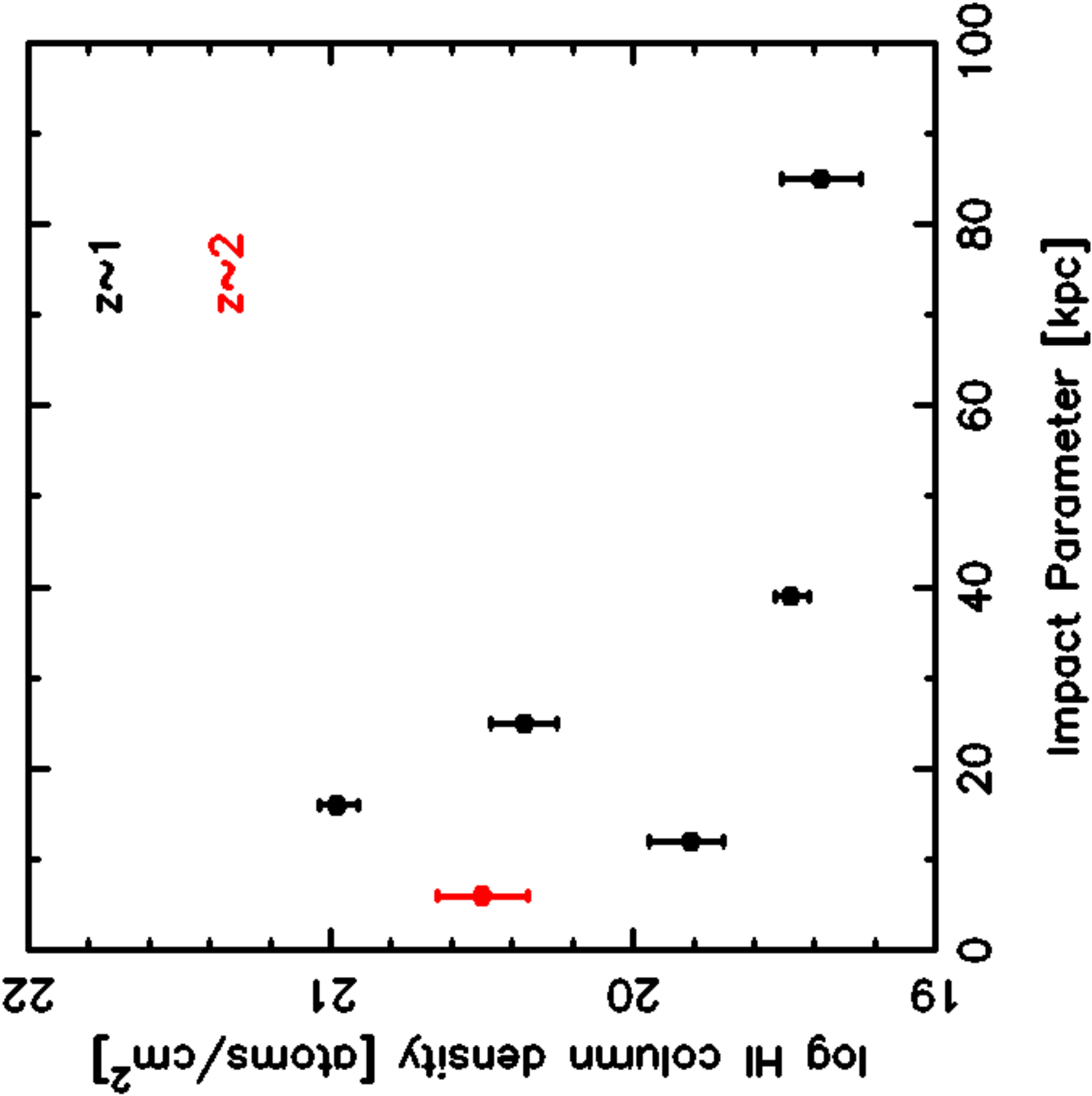}
\includegraphics[height=5.5cm, width=6cm, angle=-90]{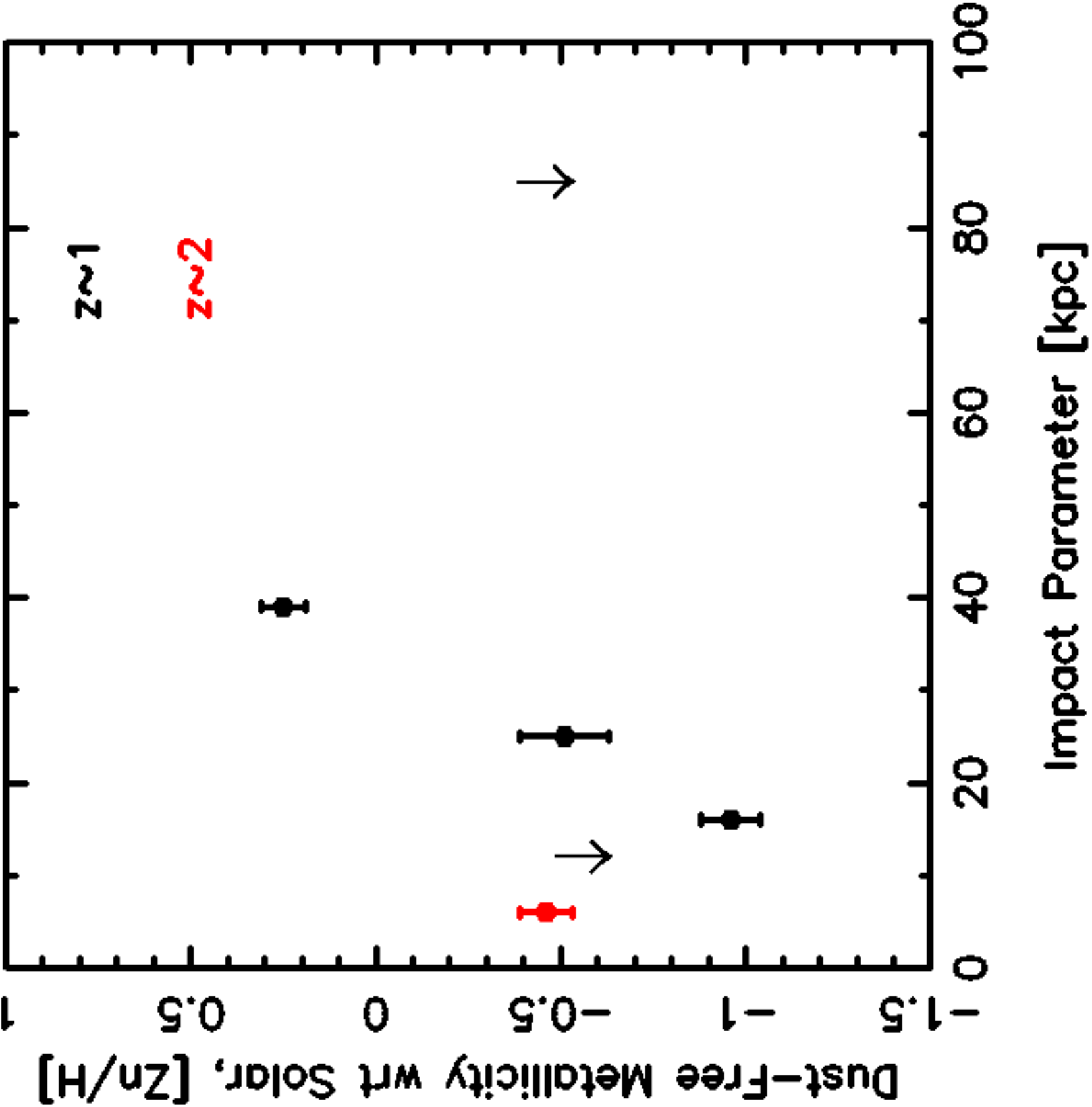}
\caption{{\bf Properties of the absorbing-galaxies detected in the SINFONI survey as a function of impact parameter.} The black points are for systems at z$\sim$1 and the red one for the system at z$\sim$2. We note that the new detection reported here has a larger impact parameter (b=86 kpc) than previously reported objects. }
\label{f:Detected}
\end{center}
\end{figure*}

Fig~\ref{f:SFR_vs_NHI} summarises the SFR estimates and limits as a function of \nhi\ for the full \nhi-selected SINFONI survey \citep[][ and this work]{peroux11a, peroux12}. The targets are selected so that they have their \nhi\ measured from HST spectroscopy. The black crosses refer to z$\sim$1 targets observed in J-band with SINFONI, while the red crosses are objects lying at z$\sim$2 observed in K-band. Taken together, the detection rates of the \nhi-selected SINFONI survey are 33\% (6/18) at z$\sim$1 and 8\% (1/12) at z$\sim$2, although number statistics remain small. Interestingly, \citet{fumagalli10} used a different technique at z$>$2 to search for absorbing-galaxies and report a significantly lower detection rate. These authors target quasar sightlines with multiple optically-thick \hi\ absorbers and use the higher-redshift system as a "blocking filter" (via its Lyman-limit absorption) to eliminate all far-ultraviolet (FUV) emission from the quasar. This allows them to directly image the rest-frame FUV continuum emission of the lower-redshift DLA, without any quasar contamination. The difference in detection rates with respect to our survey might arise from the fact that these authors measure SFR at FUV wavelengths which is prone to strong dust extinction effects although the authors have inferred minimal dust depletion from abundance studies in their sample. In our SINFONI survey, we are observing at NIR wavelengths (visible wavelengths in the absorber rest-frame) thus minimising the effect from dust obscuration. 

Even with these rather low detection rates, one might rightly raise the question whether the galaxy identified at the redshift of the absorber is directly associated with it. To answer this question, we used the \emph{REF L050N512} simulation from the OWLS project \citep{schaye10} to look at the properties of galaxies associated with absorbers. We note that this cosmological simulation successfully reproduces both the cosmic star formation rate \citep{schaye10} and \hi\ column density distribution function \citep{rahmati13} at z = 1. Following the approach presented in \cite{rahmati14}, we searched for absorbers with \nhi$>$10$^{19} $cm$^{-2}$ that have a galaxy with SFR$>$1 M$_{\odot}$/yr within a search (projected) radius of 40 kpc corresponding to our early blind search survey \citep{peroux12, peroux13}. We found that only $< 10$\% of the selected absorbers have fainter companion galaxies (with 0.01$<$SFR$<$1 M$_{\odot}$/yr) within the same search radius and at most 40\% within 100 kpc (corresponding to the targeted detection reported here). While this fraction rises strongly with increasing radius and redshift, our calculation is consistent with the idea that galaxies identified in our survey are directly associated with the DLAs and sub-DLAs.

Fig~\ref{f:SFR_vs_NHI} also shows that the SFR limits derived from non-detections are more stringent at z$\sim$1 than at z$\sim$2 and that the object detected at z$\sim$2 (towards Q2222$-$0946) has a high SFR. In addition, we target higher \nhi\ systems at high redshift than at z$\sim$1. In the sample presented here, it is interesting to point out that the non-detections of the absorbing-galaxies arise from a mixed bag of candidates based on both broad-band imaging and proximity to the quasar and/or photometric redshifts based on colours. Conversely, the SINFONI+X-Shooter detection towards SDSS J002133.27+004300.9 is based on an absorbing-galaxy candidate with no apriori spectroscopic or photometric redshift. 

\vspace{.5cm}
Fig~\ref{f:Detected} illustrates various properties of the absorbing-galaxies detected in the SINFONI survey as a function of impact parameter. Again, the black points are for systems at z$\sim$1 and the red ones for the system at z$\sim$2. We note that the new detection reported here has a larger impact parameter (b=86 kpc) than previously reported objects, although no galaxies are seen at smaller impact parameters in this field \citep{rao11}. The left panel of Fig~\ref{f:Detected} shows the SFR estimates which do not indicate a trend with impact parameter. The middle panel of Fig~\ref{f:Detected} illustrates the evolution of the \hi\ column density \nhi\ as a function of impact parameter. As expected, a correlation between these two quantities is observed in our data. This can be naturally understood in a context where these high-redshift quasar absorbers are the analogues of the \hi-rich local disks which are known to have \hi\ profiles declining with radius or where we are probing gradients in the matter density around galaxies at larger radius. Indeed, the background density close to galaxies is higher than far from them, so that larger \nhi\ quasar absorbers are present closer to over density peaks where massive galaxies are forming \citep{rahmati14}. The right panel of Fig~\ref{f:Detected} presents neutral gas \hi\ metallicity from the weakly depleted element \znii\ (Pettini et al. 1994) as a function of impact parameter. We do not see indications of decreasing metallicity with increasing distance from the centre of the galaxy. This is surprising given that metallicity gradients are known to occur in various types of galaxies \citep{swinbank12, queyrel12} and can be seen as a further evidence the gas probed in absorption might be tracing gas flows outside the galaxy's halo. We note that these measurements correspond to the absolute metallicity of the absorbing gas, independent of the emission-line metallicity of the associated galaxy.

\subsection{Gas Flow Probes}

Recent studies \citep{bordoloi11, bouche12, bouche13, schroetter15} argue that absorption aligned with the minor axis of a galaxy can be modelled by bipolar winds. Thanks to our SINFONI observations, we are able to determine the absorbing-galaxy inclination derived from the ratio of apparent major to minor axis.  In addition, we can determine the alignment between the quasar sightline and the projection on the sky of the galaxy's axes. Indeed, IFU spectroscopy is crucial to untangle the geometric effects related to the location of the quasar sight-line. It has been shown in the past by several authors \citep[e.g.][]{bouche12} that a bi-modal distribution of the azimuthal orientation of the quasar sight lines with respect to the galaxy major axis allows one to distinguish winds from gas associated with the disc. In this picture, the outflowing gas preferentially leaves the galaxy along its minor axis (path of least resistance) and inflowing gas is almost co-planar with the galaxy disc (i.e. seen preferentially along the major axis; Shen et al. 2012). The azimuthal angle between the quasar line of sight and the projected galaxy's major axis on the sky for objects detected in our SINFONI survey and from the literature are listed in Table~\ref{t:Azi}.

\begin{table*}
\begin{center}
\caption{{\bf Azimuthal angle and metallicity difference}. In our SINFONI survey, the neutral gas metallicity with respect to solar values is measured in absorption along the line-of-sight to the background quasar. The neutral gas abundances listed is based on the undepleted Zn unless stated otherwise. We note that our target selection requiring systems with known \nhi\ is key to abundance measurements in absorption. In addition, the SINFONI and X-Shooter detections of the absorbing-galaxies provide a measure of the integrated metallicity in the HII regions probed in emission. The metallicity difference between the galaxy and the absorbing gas is the underabundance in neutral phase  as defined by Lebouteiller et al. (2013): $\delta_{HI}(X)=\log (X/H)_{HII} - \log (X/H)_{HI}$.  }
\label{t:Azi}
\begin{tabular}{ccccccc}
\hline\hline
Quasar		       &Azi. Angle &$\log (X/H)_{HII}$ &$\log (X/H)_{HI}$   &$\delta_{HI}(X)$  &M$_{\rm halo}$ &References\\
&[deg]&&& &[M$_{\odot}$] &\\						                           
\hline
J002133.27+004300.9		  &10	&$<$+0.44	   &$<+0.57$               	&$<-$0.13	  &10$^{11.9}$	&This work\\
Q0302$-$223			  &70	&+0.04$\pm$0.20    &$-$0.51$\pm$0.12		&+0.55$\pm$0.23	  &--		&P\'eroux et al. (2011a)\\
Q0452$-$1640			  &26	&$-$0.46$\pm$0.20  &$-$0.96$\pm$0.08		&+0.50$\pm$0.21	  &10$^{12.8}$	&P\'eroux et al. (2012)\\
Q1009$-$0026  			  &50	&+0.04$\pm$0.80	   &+0.25$\pm$0.06		&$-$0.21$\pm$0.80 &10$^{12.6}$	&P\'eroux et al. (2011a)\\
J1422$-$0001 			  &15	&$-$1.3$\pm$0.3	&$-$0.1$\pm$0.4			&$-1.2\pm$0.5	  &10$^{11.3}$	&Bouch\'e et al (submitted)\\
J165931+373528			  &87   &$-$0.21$\pm$0.08  &$-$1.12$\pm$0.02$^a$	&$+$0.91$\pm$0.08 &-- 		&Kacprzak et al. (2014)\\
Q2222$-$0946  			  &45	&$<-$0.46	   &$-$0.46$\pm$0.07		&$<+$0.0   	  &-- 		&P\'eroux et al. (2012)\\
HE2243$-$60  			  &20	&$-$0.32$\pm$0.10  &$-$0.72$\pm$0.05		&+0.40$\pm$0.11	  &10$^{11.6}$	&Bouch\'e et al. (2013)\\
Q2352$-$0028			  &27	&$-$0.26$\pm$0.03  &$<$-0.51  			&$>+$0.25  	  &10$^{11.8}$	&P\'eroux et al. (2012)\\
\hline\hline 				       			 	 
\end{tabular}			       			 	 
\end{center}			       			 	 
\begin{minipage}{180mm}
{\bf Note:} \\
{\bf $^a$:} [X/H] based on \mgi, \mgii, \siii, \siiii\ and CLOUDY modelling \citep{kacprzak14}.\\
\end{minipage}
\end{table*}

Besides these geometry aspects, a key diagnostic to untangle infalling from outflowing gas around galaxies is the metallicity of the gas. Indeed, metals are known to be produced by stars within galaxies, which will then further enrich the outflowing gas. Conversely, infalling gas feeding galaxies from the filaments of the cosmic web is expected to be pristine. Using these lines of argument, \citet{lehner13} and Wotta et al. (in preparation), report a bimodality in the metallicity distribution of Lyman Limit Systems (LLS) at z$<$1. These authors interpret this result as a signature of some LLS probing infalling gas while other LLS would be related to the absorbing-galaxy or associated outflowing gas. A similar behaviour is reported by Quiret et al. (submitted) for sub-DLAs with z$<$1.25. We note however that these measurements do not provide information about the metallicity of the absorbing-galaxy. 

In our SINFONI survey, the neutral gas metallicity with respect to solar values are measured in absorption along the line-of-sight to the background quasar. A summary of these measurements is included in Table~\ref{t:Azi}. The neutral gas abundance is derived from the undepleted Zn abundance in all cases. We note that our target selection including systems with known \nhi\ is key to abundance measurements in absorption. In addition, the SINFONI and X-Shooter detections of the absorbing-galaxies provide a measure of the integrated metallicity in the HII regions probed in emission. The error estimates in these metallicity differences are based on the errors in the emission and absorption metalicities added in quadrature. Again, these values are listed in Table~\ref{t:Azi} with respect to solar. 

Given this information, we are able to compute the metallicity difference between the stars in the galaxy and the absorbing gas in the SINFONI sample. We use the underabundance in neutral phase as defined by \citet{lebouteiller13}: 

\begin{equation}
\delta_{HI}(X)=\log (X/H)_{HII} - \log (X/H)_{HI}
\end{equation}

These values provide an estimate of the relative metallicity of the gas probed in absorption with respect to the metallicity in the absorbing-galaxy itself. Here, we use neutral gas abundances based on the element Zn, which is only weakly depleted onto dust grains \citep{pettini94}.  A positive $\delta_{HI}(X)$ value indicates a neutral gas metallicity lower than the galaxy's HII regions, a possible indicator of gas inflow. Conversely, a null or negative $\delta_{HI}(X)$ value indicates a equal or higher metallicity in the absorbing gas than in the galaxy, possibly indicating outflowing gas enriched by star formation within the galaxy.

\begin{figure*}
\begin{center}
\includegraphics[height=8.cm, width=11.cm, angle=0]{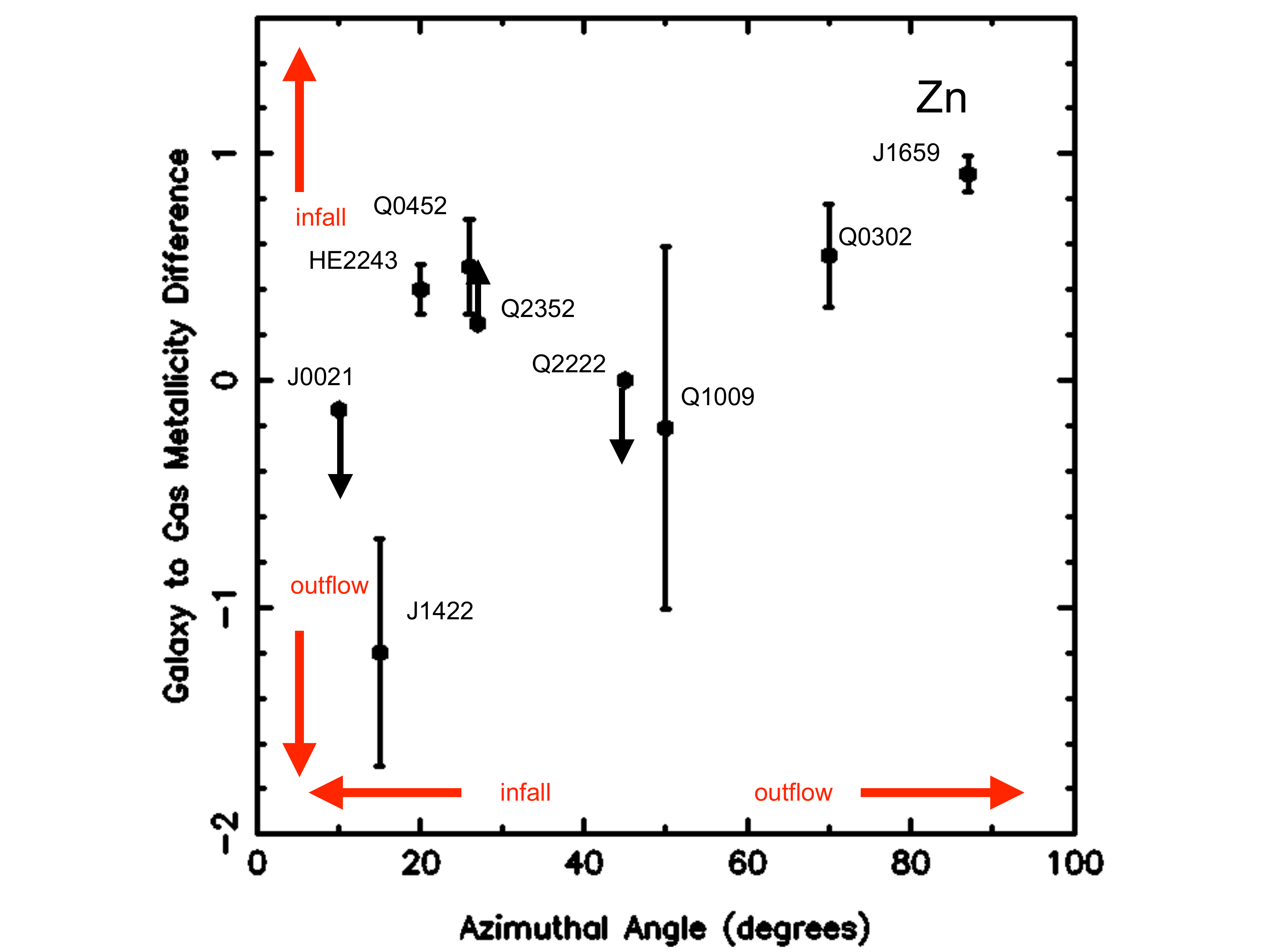}
\caption{{\bf Metallicity difference as a function of azimuthal angle.} In this plot, we expect \citep{vandevoort12} infalling gas to lie in the upper left corner (low azimuthal angle and high $\delta_{HI}(X)$ value), while outflowing gas would lie in the lower right corner (high azimuthal angle and zero or negative $\delta_{HI}(X)$ value). These are indicated as red arrows in the figure. Additional observations are necessary to better relate metallicity and geometry in gas flows.}
\label{f:Azi}
\end{center}
\end{figure*}

In Fig~\ref{f:Azi}, we plot the metallicity difference as a function of azimuthal angle, two independent indicators of gas flows. In this plot, we expect infalling gas to lie in the upper left corner (low azimuthal angle and high $\delta_{HI}(X)$ value), while outflowing gas would lie in the lower right corner (high azimuthal angle and low $\delta_{HI}(X)$ value). These are indicated as red arrows in the figure. In addition, we plot the 8 objects from our SINFONI survey as well as one object from the literature \citep{kacprzak14} for which all these measurements are available. We note that the few observations available so far do not correlate as expected from simple assumptions. Indeed, the galaxy to gas metallicity difference appears positive in systems with high azimuthal angles while low azimuthal angles cover various metallicity differences including negative values. Quantitatively, we derive a correlation coefficient of 0.30 indicating a weak anti-correlation and a probability P=0.43 of no-correlation. In particular, we report no system which would be equally or more metal-rich in the distant absorbing gas than in the emission line gas in the galaxy and aligned with the minor axis of the disk as expected from outflows (i.e. bottom right corner of Fig~\ref{f:Azi}).

\subsection{Discussion}

While we chose to present here neutral gas phase abundances derived from Zn, we note that using Fe provides both a similar number of measurements and a similar trend, only shifted upwards in metallicity values and with additional complications due to dust depletion. Since only a few Si measurements are available for the systems in this sample, and Si is anyway depleted onto dust grains, no robust conclusions can be derived based on this element. It is important to note that the metallicity estimates measured in the galaxy and in its outskirts are based on different phases of gas: the emission metallicity traces the ionized gas, while the absorption metallicity traces the neutral phase of the gas. In addition, the emission metallicities reported here use a number of different indicators ranging from R$_{\rm 23}$ \citep{pagel79} to the \nii/H$-\alpha$ parameter \citep{Perez09}. For example, 5 systems in our sample (namely J002133.27+004300.9, Q1009$-$0026, J1422$-$0001, J165931+373528 and Q2222$-$0946) have their HII metallicity derived from \nii/H$-\alpha$, including two (J002133.27+004300.9 and Q2222$-$0946) for which \nii\ was not detected giving rise to an upper limit in metallicity. The uncertainties in the measurements for the remaining three systems are evidenced by fairly large error bars. The other objects have more reliable \hii\ metallicity measured from R$_{\rm 23}$ based on multi-emission lines detections in X-Shooter observations \citep{peroux11a, peroux12, bouche13}. 

Similarly, the robustness of the measurements of the azimuthal angle in these systems varies from one object to another. For example, the kinematic properties of the system presented here, J002133.27+004300.9, is rather poorly constrained as described in previous sections. Another example of this is Q2222$-$0946, which is a compact galaxy with a small impact parameter to the quasar's line-of-sight (b=6 kpc). \cite{peroux12} derive an azimuthal angle of 45 degrees which on its own cannot be used to disentangle inflows from galactic winds based on geometric arguments. 

Finally, we note that the system reported here, J002133.27+004300.9, has a large impact parameter (b=85 kpc) which might be uncommon for DLA-galaxy association (Rahmani et al. in prep), but typical of CGM regions \citep{Shull14b}. On the other hand, some other systems have significantly smaller impact parameters (namely Q2222$-$0946 again) which questions whether the gas probed in absorption is related to CGM regions or more likely to internal parts of the galaxy.  

Keeping these limitations in mind, it is interesting to explore the interpretation of this rather unexpected result. Clearly, Fig~\ref{f:Azi} suggests that the phenomenon of gas flows may be more complicated and metallicity and geometry might be rather indirect tracers of gas flows \citep{lehner13, bouche07}. For example, not all of the gas detected in absorption might be related to the CGM of the galaxy. Indeed, some of it could be related to the rotating gas in the disk of the galaxy itself \citep{bouche13}. In fact, a comparison of the V$_{\rm max}$ values when available and the total absorption width might bring some quantitative information on this effect, but we note that in most cases it will bring the data points to higher $\delta_{HI}(X)$ values (because V$_{\rm max}$ represent just a fraction of the total absorption profile in most cases), whether they lie at small or large azimuthal angles. Specifically, the V$_{\rm max}$ in the case of Q2222$-$0946 and J165931+373528 is unknown because the object is compact in the former case and no IFU data are available in the latter case. Interestingly, the whole of the absorption profile along Q1009$-$0026 matches quite well the V$_{\rm max}$ in that system, so that this data point is not expected to move towards more positive $\delta_{HI}(X)$ values. In the case of J1422$-$0001, Bouch\'e et al. (submitted) report that the inflowing material is considerably metal rich hinting towards recycling material, hence its position in the lower left corner of the diagram. For the remaining systems V$_{\rm max}$ is smaller that the actual spread of the absorption profile, and as such would all have slightly more positive $\delta_{HI}(X)$ values. Also, Fig~\ref{f:Azi} suggests that some systems (in the upper right corner) may consist of outflows of metal-poor gas, possibly hinting to gas gets ejected before undergoing much star formation.

In addition, simulations indicate that inflows and outflows are complex phenomena. Hot accretion inflows are not expected to aligned along any preferred direction \citep{ocvirk08, nelson15}. Indeed, some of our galaxies have high halo masses (see Table~\ref{t:Azi}), and may be cases of hot mode accretion. Gas flows can potentially be affected by numerous shocks \citep{keres05}. Similarly, some of the gas is likely to be multi-phase and clumpy therefore limiting the 
extent to which a given absorption region represents the halo, given that it comes from only one quasar on a random sightline. Moreover, some of the gas expelled by outflows might not escape the potential well of the galaxy, falling back as galactic fountains but already polluted by metals from star formation \citep{dekel86, efstathiou00, cresci15}. Such systems will be as metal-rich in absorption as the galaxy (lower part of Fig~\ref{f:Azi}) and potentially still be co-planar to the disk (left part of Fig~\ref{f:Azi}). Such processes might therefore explain the objects populating the bottom left corner of the figure. An example of this is the absorbing system towards J1422$-$0001. Overall, additional observations are necessary to better relate metallicity and geometry in gas flows. With the new MUSE instrument on VLT (Contini et al. 2015) and on longer-term the HARMONI IFU on the E-ELT \citep{evans15}, prospects to make progress in this field are good. 

\section{Conclusion}

We report new observations with SINFONI and X-Shooter of absorbing-galaxy candidates at z$\sim$1. The targets have been selected to have a known sky position and an identification based on a proximity argument or based on a photometric or a spectroscopic redshifts. We emphasize that this is not a blind search, but a set of eight z$\sim$1 quasar/absorbing-galaxy matched pairs identified by others. Our findings can be summarised as follows:

\begin{itemize}

\item We report the detection with SINFONI of the \ha\ emission line of one sub-DLA at \zabs=0.94187 with \lognhi=19.38$^{+0.10}_{-0.15}$ towards SDSS J002133.27+004300.9 (out of 6 targets observed). We estimate the SFR to be SFR=3.6$\pm$2.2 M$_{\odot}$/yr in that system. A detailed kinematic study indicates a maximum circular velocity V$_{\rm max}$=80$\pm$40 km/s with a dispersion $\sigma$=123$\pm$11 km/s. Based on these, we measure a dynamical mass M$_{\rm dyn}$=10$^{9.9\pm0.4}$ M$_{\odot}$ and a halo mass M$_{\rm halo}$=10$^{11.9\pm0.5}$ M$_{\odot}$. The remaining targets are not detected in our SINFONI observations.

\item We report the detection with X-Shooter of the system above as well as one emission line at the redshift \zabs=0.7402 of another DLA with \lognhi=20.4$\pm$0.1 toward Q0052$+$0041. The \ha\ flux in the former system is in perfect agreement with the value derived from the SINFONI observations. The latter system is not detected in \ha\ but in \oii\ as previously reported in the literature from Keck/ESI spectroscopy (Lacy et al. 2003). Based on the detected \oii\ flux, we estimate SFR=5.3 $\pm$0.7M$_{\odot}$/yr in the absorbing-galaxy towards Q0052$+$0041.

\item We report the detection of another three objects with X-Shooter at the expected position of the absorbing-galaxies in the field of SDSS J011615.51-004334.7, SDSS J122836.8+101841.7 and SDSS J152102.00-000903.1. While these objects are detected in the continuum at NIR wavelengths, a careful visual inspection of the 2D images does not reveal any predominant emission lines for the two latter. The nature and redshift of these objects are therefore not constrained. The former object presents a power-law continuum and an emission line consistent with the redshift of the background quasar, but this could be due in part to contamination from the nearby quasar.

\item In two cases (in the field of J045647.17+040052.9 and Q0826-2230), no objects are detected at the expected position of the absorbing-galaxies. We note that both these quasars are two order of magnitudes brighter than the other targets in the sample (V Mag$\sim$16), thus producing a high contrast between the quasar and the faint absorbing-galaxy. The former system is also characterised by a small angular separation (impact parameter b=0.8"). 
 
\item For all the objects detected in our five-year SINFONI survey, we compute the metallicity difference between the galaxy and the absorbing gas using the underabundance in neutral phase  as defined by \citet{lebouteiller13}: $\delta_{HI}(X)=\log (X/H)_{HII} - \log (X/H)_{HI}$. We compare this quantity with the azimuthal angle in the same objects to relate these two independent indicators of gas flows. We find that these quantities do not correlate as expected from simple assumptions. Indeed, some of the gas might fall back on the galaxy as enriched infalling gas. Moreover, multi-phase or clumpy gas might also complicate the simple picture. Additional observations are necessary to relate these two independent probes of gas flows around galaxies. 
\end{itemize}

\section*{Acknowledgements}
We thank Nicolas Bouch\'e, Max Pettini and Crystal Martin for inputs on the work presented here. We also thank the anonymous referee for many suggestions that have improved the paper. We would like to thank the Paranal and Garching staff at ESO for performing the observations in service mode and the instrument team for making a reliable instrument. CP thanks the ESO science visitor program for support. SQ thanks CNRS and CNES (Centre National d'Etudes Spatiales) for support for his PhD. VPK acknowledges partial support from the U.S. National Science Foundation grant AST/1108830 and NASA grant  NNX14AG74G. Additional support from a NASA/STScI grant for program 13733 is gratefully acknowledged. This work has been carried out thanks to the support of the OCEVU Labex (ANR-11-LABX-0060) and the A*MIDEX project (ANR-11-IDEX-0001-02) funded by the "Investissements d'Avenir" French government program managed by the ANR.

\bibliographystyle{mn2e}
\bibliography{/Users/celine/Astro/Paper/SINFOVI/bibliography.bib}

\label{lastpage}
\end{document}